\documentclass[11pt]{article}
\usepackage{amssymb,amsmath,amsfonts}
\usepackage{graphicx}
\usepackage{graphics}
\usepackage{eepic,epsfig}

\@addtoreset{equation}{section}

\makeatother

\begin{document}

\title{Cosmic string and brane induced effects on the fermionic \\
vacuum in AdS spacetime}
\author{S. Bellucci$^{1}$\thanks{%
E-mail: bellucci@lnf.infn.it },\thinspace\ W. Oliveira dos Santos$^{1,2}$%
\thanks{%
E-mail: wagner.physics@gmail.com},\thinspace\ E. R. Bezerra de Mello$^{2}$%
\thanks{%
E-mail: emello@fisica.ufpb.br},\thinspace\ A. A. Saharian$^{3}$\thanks{%
E-mail: saharian@ysu.am} \\
\\
\textit{$^1$ INFN, Laboratori Nazionali di Frascati,}\\
\textit{Via Enrico Fermi 54, 00044 Frascati, Italy} \vspace{0.3cm}\\
$^{\mathit{2}}$\textit{Departamento de F\'{\i}sica, Universidade Federal da
Para\'{\i}ba}\\
\textit{58.059-970, Caixa Postal 5.008, Jo\~{a}o Pessoa, PB, Brazil}\vspace{%
0.3cm}\\
\textit{$^{3}$Department of Physics, Yerevan State University,}\\
\textit{1 Alex Manoogian Street, 0025 Yerevan, Armenia}}
\maketitle

\begin{abstract}
We investigate the combined effects of a magnetic flux-carrying cosmic
string and a brane on the fermionic condensate (FC) and on the vacuum
expectation value (VEV) of the energy-momentum tensor for a massive charged
fermionic field in background of 5-dimensional anti-de Sitter (AdS)
spacetime. The brane is parallel to the AdS boundary and it divides the
space into two regions with distinct properties of the fermionic vacuum. For
two types of boundary conditions on the field operator and for the fields
realizing two inequivalent representations of the Clifford algebra, the
brane-induced contributions in VEVs are explicitly separated. The VEVs are
even periodic functions of the magnetic flux, confined in the core, with the
period of flux quantum. Near the horizon the FC and the vacuum
energy-momentum tensor are dominated by the brane-free contribution, whereas
the brane-induced part dominates in the region near the brane. Both the
contributions vanish on the AdS boundary. At large distances from the cosmic
string, the topological contributions in the VEVs, as functions of the
proper distance, exhibit an inverse power-law decrease in the region between
the brane and AdS horizon and an exponential decrease in the region between
the brane and AdS boundary. We show that the FC and the vacuum energy
density can be either positive or negative, depending on the distance from
the brane. Applications are discussed in fermionic models invariant under
the charge conjugation and parity transformation and also in $Z_{2}$%
-symmetric braneworld models. By the limiting transition we derive the
expressions of the FC and the vacuum energy-momentum tensor for a cosmic
string on 5-dimensional Minkowski bulk in the presence of a boundary
perpendicular to the string.
\end{abstract}


\bigskip

\section{Introduction}

\label{Int}

The vacuum polarization is among the most interesting manifestations of the
nontrivial properties of quantum vacuum. Various types of sources for the
polarization of vacuum have been considered in the literature. They include
external electromagnetic and gravitational fields or boundary conditions on
quantum fields induced by the presence of boundaries having different
physical nature. Examples of the latter are the material boundaries in
quantum electrodynamics, interfaces separating different phases of physical
systems, horizons, etc. The spectrum of vacuum fluctuations is influenced by
the boundary conditions and, as a consequence, the expectation values of
physical observables are shifted. Those modifications depend on the bulk and
boundary geometries, on the boundary conditions imposed and are known as the
Casimir effect (for reviews see \cite{Eliz94C}). Another interesting source
for the vacuum polarization is the nontrivial topology induced by
compactification of spatial dimensions or by the presence of topological
defects. The corresponding effects play an important role in both
high-energy quantum field theoretical models and in effective models
describing the condensed matter systems. In the present paper we consider an
exactly solvable problem for the polarization of the fermionic vacuum under
the combined influence of several sources: background gravitational field, a
topological defect of the cosmic string type and a boundary.

As to the background geometry we will take a locally anti-de Sitter (AdS)
spacetime. There are several motivations for that. The AdS spacetime is the
maximally symmetric solution of the Einstein equations with the negative
cosmological constant as the only source for the gravitational field. The
high degree of symmetry allows to obtain closed analytic expressions for
local characteristics of the vacuum state. That is important for elucidating
the features of the influence of the gravitational field on the properties
of quantum field-theoretical systems in more complicated geometries. Of
course, the results obtained for the AdS bulk are also of separate interest,
due to its important role as the ground state in supergravity and
superstring theories and also in two exciting developments of the
contemporary theoretical physics, namely, in braneworld models with large
extra dimensions \cite{Maar10} and in the AdS/conformal field theory (CFT)
correspondence \cite{Ahar00}. The braneworlds provide a geometrical solution
of the hierarchy problem for the energy scales in particle physics and have
been extensively used in cosmological and astrophysical contexts. The
AdS/CFT correspondence is a realization of the holographic type duality
between supergravity or string theories on background of the AdS spacetime
and CFT located on its boundary. It is an interesting way to investigate
non-perturbative effects in the theory by mapping them onto perturbative
effects in the dual theory and has important applications in high-energy
physics and in a number of condensed matter systems.

In the consideration below the nontrivial topology is induced by the
presence of a two-dimensional topological defect, carrying a gauge field
flux along its core. The investigation will be done within the framework of
a simple model that reduces the influence of the defect on the geometry to
the generation of a planar angle deficit in the spacetime line element. The
image of the defect on the boundary of the AdS spacetime corresponds to the
standard idealized cosmic string geometry in (3+1)-dimensional spacetime and
we use the term cosmic string for the defect under consideration as well.
The cosmic strings form a class of topological defects created as a result
of symmetry breaking phase transitions during the expansion in the early
Universe (see \cite{Vile94} for reviews). Among other types of topological
defects (domain walls, monopoles, textures), they have been most frequently
considered in the literature. That is related to interesting physical
effects induced by the presence of cosmic strings, such as the generation of
gamma ray bursts, high-energy cosmic rays, gravitational waves and scale
invariant cosmological perturbations. The formation of macroscopic defects
of cosmic string type has been also predicted within the framework of
fundamental string theory \cite{Witt85}. An example is provided by the
brane-inflation scenario where cosmic strings are produced towards the end
of inflation. Cosmic string type solutions on AdS bulk and in braneworld
models have been considered, for example, in \cite{Dehg01,Davi01}. The
polarization of quantum vacuum around cosmic strings has been mainly
considered for the background Minkowski geometry (see, for instance,
references given in \cite{Bell14,Mota18}). The combined effects of the
gravitational field and of a cosmic string on the local characteristics of
scalar, fermionic and electromagnetic vacua in the case of the de Sitter
(dS) background were discussed in \cite{Beze09}. Another exactly solvable
background with a cosmic string corresponds to the AdS bulk. The vacuum
expectation values (VEVs) of the field squared and of the energy-momentum
tensor induced by a straight cosmic string on that bulk have been
investigated in \cite{Beze12} for scalar and fermionic fields.

As an additional source of the vacuum polarization we will consider a brane
parallel to the AdS boundary with two types of boundary conditions on the
fermionic field operator. The first one corresponds to the bag boundary
condition and the second one differs by the sign of the term containing the
normal to the brane. In braneworld models of the Randall-Sundrum type those
boundary conditions are dictated by the $Z_{2}$-symmetry. The boundary
conditions give rise to Casimir type contributions in the VEVs of physical
observables. We investigate those contributions for the fermionic condensate
(FC) and for the expectation value of the energy-momentum tensor for fields
realizing two inequivalent irreducible representations of the Clifford
algebra. The corresponding VEV of the current density for a charged
fermionic field has been considered in \cite{Bell21}. Another type of
conditions imposed on the operator of a quantum field appears in models with
compact spatial dimensions. Similar to the case of boundary conditions, the
periodicity conditions along compact dimensions give rise to topological
Casimir contributions in the vacuum characteristics. In a locally AdS
spacetime with compact dimensions the combined effects of branes and
compactification on the vacuum energy and on the VEV of the energy-momentum
tensor have been investigated in \cite{Flac03}. The corresponding vacuum
currents for charged scalar and fermionic fields were discussed in \cite%
{Beze15,Bell18}. An additional topological influence of a cosmic string on
the vacuum currents has been considered in \cite{Oliv19,Bell20}. The FC and
the VEV of the energy-momentum tensor for a charged massive fermionic field
were investigated in \cite{Bell21b,Bell22}. The influence of the brane on
those characteristics of the fermionic vacuum in the geometry of a cosmic
string on a locally AdS bulk is the main concern in the present paper. The
choice of the FC and the energy-momentum tensor is motivated by their
important role in the self-consistent description of the dynamics of the
system including the influence of the gravitational field. Being the VEVs of
local operators, because of the global nature of the vacuum state, they also
contain information on the global characteristics of the background
geometry. In particular, the FC is the central quantity in the discussions
of dynamical breaking of chiral symmetry (see, for example, \cite{Eliz94}
for the chiral symmetry breaking in Nambu-Jona-Lasino and Gross-Neveu models
on background of a curved spacetime with non-trivial topology).

The paper is organized as follows. In Section \ref{sec1} we describe the
system under consideration and present a complete set of fermionic modes in
both the regions of the spacetime separated by a brane. By using those
modes, in Section \ref{sec:FCond} the FC is investigated. A general formula
is provided with explicit separation of the brane-induced contribution.
Various limiting cases and the behavior of the FC in asymptotic regions of
the parameters are discussed. Similar analysis for the VEV of the
energy-momentum tensor is presented in Section \ref{sec:EMT0}. The model
under consideration lives on an odd dimensional spacetime and the
corresponding Clifford algebra has two inequivalent irreducible
representations. In Section \ref{sec:Reps} we discuss the relations of the
fields realizing those representations with the field discussed in the
previous sections. Depending on the boundary conditions imposed on the
separate fields, in addition to the bag boundary condition, one needs to
consider the boundary condition that differs by the sign of the term
involving the normal to the brane. The corresponding vacuum densities are
presented. We also discuss the VEVs in models invariant under the charge
conjugation and parity transformation. Applications are given in $Z_{2}$%
-symmetric braneworld models of the Randall-Sundrum type with a single brane
and in the presence of a cosmic string. The main results of the paper are
summarized in Section \ref{Conc}. In Appendix \ref{sec:App1} we present the
details of the separation of the brane-induced contributions in the FC and
in the VEV of the energy-momentum tensor.

\section{Background geometry and fermionic modes}

\label{sec1}

In this section we present the background geometry of a 5-dimensional
locally AdS spacetime in the presence of an idealized cosmic string and the
complete set of the positive and negative energy fermionic modes, being the
solutions of the Dirac equation subject to the bag boundary condition on a
brane parallel to the AdS boundary.

We consider a (4+1)-dimensional spacetime covered by the spatial coordinates
$(r,\phi ,y,z)$, with $r\geq 0$, $0\leq \phi \leq 2\pi /q$, $-\infty
<y,z<+\infty $. The local geometry is described by the line element
\begin{equation}
ds^{2}=e^{-2y/a}\left( dt^{2}-dr^{2}-r^{2}d\phi ^{2}-dz^{2}\right) -dy^{2}\ ,
\label{ds1}
\end{equation}%
where $t\in (-\infty ,\ \infty )$. In the special case $q=1$, it corresponds
to 5-dimensional AdS spacetime generated by the cosmological constant $%
\Lambda =-6/a^{2}$. For $q\neq 1$, the line element (\ref{ds1}) describes an
idealized topological defect with the core localized on the two-dimensional
spatial hypersurface $r=0$. The geometry of the core is given by the element
$ds_{\mathrm{core}}^{2}=e^{-2y/a}\left( dt^{2}-dz^{2}\right) -dy^{2}$. Note
that the presence of the defect does not change the local characteristics of
the geometry outside the core and they coincide with those for
(4+1)-dimensional AdS spacetime. Two projections of the element (\ref{ds1})
describe idealized cosmic strings with the planar angle deficit $2\pi
(1-1/q) $. The first one, corresponding to a hypersurface $y=\mathrm{const}$%
, presents a cosmic string in (3+1)-dimensional Minkowski spacetime. In
particular, that is the case for the geometry of the conformal field theory
in the context of the AdS/CFT correspondence. The second projection with $z=%
\mathrm{const}$ corresponds to a cosmic string in background of
(3+1)-dimensional AdS spacetime. In the discussion below it will be
convenient to work in the coordinates where the geometry outside the core $%
r=0$ is conformally flat. Introducing a new coordinate $w$ in accordance
with $w=ae^{y/a}$, the line element is written in the form conformally
related to a 5-dimensional Minkowski spacetime:
\begin{equation}
ds^{2}=g_{\mu \nu }dx^{\mu }dx^{\nu }=\left( \frac{a}{w}\right) ^{2}\left(
dt^{2}-dr^{2}-r^{2}d\phi ^{2}-dw^{2}-dz^{2}\right) \ ,  \label{ds2}
\end{equation}%
with $x^{\mu }=(t,r,\phi ,w,z)$. This explicitly shows the conformal
connection to the geometry of a cosmic string in background of
(4+1)-dimensional Minkowski spacetime with the line element
\begin{equation}
ds_{\mathrm{M}}^{2}=dt^{2}-dr^{2}-r^{2}d\phi ^{2}-dw^{2}-dz^{2},
\label{ds2M}
\end{equation}%
where, as before, $0\leq \phi \leq 2\pi /q$. The new coordinate is defined
in the interval $0\leq w<\ \infty $ and the endpoints $w=0$ and $w=\infty $
correspond to the AdS boundary and horizon, respectively.

Having specified the background geometry we turn to the field content. A
quantum fermionic field $\psi $ will be considered that realizes an
irreducible representation of the Clifford algebra for the Dirac matrices $%
\gamma ^{\mu }$. In a 5-dimensional spacetime it is presented by a
4-component spinor. The Dirac matrices are expressed in terms of the
corresponding flat spacetime matrices $\gamma ^{(b)}$ as $\gamma ^{\mu
}=e_{(b)}^{\mu }\gamma ^{(b)}$, where $e_{(b)}^{\mu }$ are the vielbein
fields. The dynamics of the field is described by the Lagrangian density
\begin{equation}
L=\bar{\psi}\left( i\gamma ^{\mu }\mathcal{D}_{\mu }-sm\right) \psi \ ,
\label{Lsp}
\end{equation}%
with the covariant derivative operator $\mathcal{D}_{\mu }=\partial _{\mu
}+\Gamma _{\mu }+ieA_{\mu }$ and the Dirac adjoint $\bar{\psi}=\psi
^{\dagger }\gamma ^{(0)}$. Here, $\Gamma _{\mu }$ is the spin connection, $%
A_{\mu }$ is the vector potential for a gauge field and $\pm e$ are the
charges of the field quanta. In odd dimensional spacetimes the Clifford
algebra has two inequivalent irreducible representations and the values $s=1$
and $s=-1$ of the parameter $s$ distinguish two fields realizing those
representations (see the discussion in Section \ref{sec:Reps}). Taking the
vielbein fields $e_{(b)}^{\mu }=(w/a)\delta _{b}^{\mu }$, we use the
following representation for the $4\times 4$ Dirac matrices:
\begin{equation}
\gamma ^{0}=\frac{iw}{a}\left( {%
\begin{array}{cc}
0 & -1 \\
1 & 0%
\end{array}%
}\right) \ ,\;\gamma ^{l}=\frac{iw}{a}\left( {%
\begin{array}{cc}
-\sigma ^{l} & 0 \\
0 & \sigma ^{l}%
\end{array}%
}\right) \ ,\;\gamma ^{4}=\frac{iw}{a}\left( {%
\begin{array}{cc}
0 & 1 \\
1 & 0%
\end{array}%
}\right) ,  \label{gam}
\end{equation}%
where the $2\times 2$ Pauli matrices $\sigma ^{l}$, $l=1,2,3$, corresponding
to the coordinates $(r,\phi ,w)$, are given by
\begin{equation}
\sigma ^{1}=\left( {%
\begin{array}{cc}
0 & e^{-iq\phi } \\
e^{iq\phi } & 0%
\end{array}%
}\right) \ ,\ \sigma ^{2}=\frac{i}{r}\left( {%
\begin{array}{cc}
0 & -e^{-iq\phi } \\
e^{iq\phi } & 0%
\end{array}%
}\right) \ ,  \label{Pauli}
\end{equation}%
and $\sigma ^{3}=\mathrm{diag}(1,-1)$.

The equation of motion corresponding to the Lagrangian density (\ref{Lsp})
reads
\begin{equation}
\left( i\gamma ^{\mu }\mathcal{D}_{\mu }-sm\right) \psi =0\ .
\label{DiracEq}
\end{equation}%
In the discussion below we assume a special configuration of a classical
gauge field with the vector potential $A_{\mu }=(0,0,A,0,0)$ having the only
nonzero constant covariant component $A_{2}=A$. This corresponds to a
magnetic flux $\Phi =-2\pi A/q$, running along the core of the defect.
Outside the core the field tensor $F_{\mu \nu }=\partial _{\mu }A_{\nu
}-\partial _{\nu }A_{\mu }$ vanishes and the effect of the magnetic flux on
the local characteristics of the fermionic vacuum in the region $r>0$ is
purely topological. Here we are interested in the effects of a codimension
one brane, parallel to the AdS boundary, on the vacuum FC and on the VEV of
the energy-momentum tensor. Assuming that the brane is located at $w=w_{0}$,
the MIT bag boundary condition will be imposed on the field operator:
\begin{equation}
(1+i\gamma ^{\mu }n_{\mu })\psi =0\ ,\ \ w=w_{0}\ ,  \label{MITbc}
\end{equation}%
where $n_{\mu }$ is the normal to the brane. The latter is given as $n_{\mu
}=\delta _{\mu }^{3}a/w$ in the region $0\leq w\leq w_{0}$ (referred to as
the L(left)-region) and as $n_{\mu }=-\delta _{\mu }^{3}a/w$ for the region $%
w_{0}\leq w<\infty $ (R(right)-region). The value of the $y$-coordintae
corresponding to the location of the brane we will denote by $y_{0}$.

The analysis of the FC and of the mean energy-momentum tensor on the pure
AdS spacetime considering the presence of branes has been developed in \cite%
{Eliz13}. As to the investigation of those VEVs on the AdS background in the
presence of a magnetic flux-carrying cosmic string, it was considered in
\cite{Bell21b,Bell22}. Here in this paper, we want to investigate the
influence of the gravitational field, the nontrivial spatial topology and
the presence of the brane on the local properties of the FC and the VEV of
the energy-momentum tensor.

The VEVs of physical observables bilinear in the field operator can be
expressed in terms of the mode sum over the complete set of solutions of the
field equation. For the system under consideration, those solutions were
obtained in \cite{Bell21}. They are specified by the set of quantum numbers $%
\sigma =(\lambda ,j,p,k_{z},\eta )$, where $0\leq \lambda ,p<\infty $, $%
j=\pm 1/2,\pm 3/2,\ldots $, $-\infty <k_{z}<+\infty $, and $\eta =\pm 1$.
The energy is expressed as%
\begin{equation}
E=\sqrt{\lambda ^{2}+p^{2}+k_{z}^{2}}.  \label{E}
\end{equation}%
The positive (upper sign) and negative (lower sign) energy fermionic mode
functions are presented as
\begin{equation}
\psi _{\sigma }^{(\pm )}(x)=C_{\sigma }^{(\pm )}w^{5/2}\left( {%
\begin{array}{c}
J_{\beta _{j}}(\lambda r)W_{\nu _{1}}(pw)e^{-iq\phi /2} \\
\mp s\epsilon _{j}\kappa _{\eta }b_{\eta }^{(\pm )}J_{\beta _{j}+\epsilon
_{j}}(\lambda r)W_{\nu _{2}}(pw)e^{iq\phi /2} \\
is\kappa _{\eta }J_{\beta _{j}}(\lambda r)W_{\nu _{2}}(pw)e^{-iq\phi /2} \\
\pm i\epsilon _{j}b_{\eta }^{(\pm )}J_{\beta _{j}+\epsilon _{j}}(\lambda
r)W_{\nu _{1}}(pw)e^{iq\phi /2}%
\end{array}%
}\right) e^{iqj\phi +ik_{z}z\mp iEt}\ ,  \label{FermMod}
\end{equation}%
where $J_{\beta _{j}}(\lambda r)$ is the Bessel function \cite{Abra} of the
order
\begin{equation}
\beta _{j}=q|j+\alpha |-\epsilon _{j}/2\ ,\;\epsilon _{j}=\mathrm{sgn}%
(j+\alpha ),  \label{betj}
\end{equation}%
and the parameter $\alpha $ is expressed in terms of the magnetic flux as $%
\alpha =-\Phi /\Phi _{0}$, being $\Phi _{0}=2\pi /e$ the quantum flux. In
the coefficients of the spinor components we have introduced the notations
\begin{eqnarray}
\kappa _{\eta } &=&\frac{-k_{z}+\eta \sqrt{k_{z}^{2}+p^{2}}}{p}\ ,  \notag \\
b_{\eta }^{(\pm )} &=&\frac{E\mp \eta \sqrt{k_{z}^{2}+p^{2}}}{\lambda },
\label{bs}
\end{eqnarray}%
with $b_{\eta }^{(+)}b_{\eta }^{(-)}=1$. The dependence of the fermionic
modes on the coordinate $w$ enters in the form of the functions $%
w^{5/2}W_{\nu _{1}}(pw)$ and $w^{5/2}W_{\nu _{2}}(pw)$, where%
\begin{equation}
W_{\nu }(pw)=C_{1}J_{\nu }(pw)+C_{2}Y_{\nu }(pw)  \label{Wn1}
\end{equation}%
is a linear combination of the Bessel and Neumann functions and
\begin{equation}
\nu _{l}=ma+(-1)^{l}s/2.  \label{n12}
\end{equation}%
The coefficients in the linear combination depend on the region of the space
under consideration. In the L-region we take $(C_{1},C_{2})=(1,0)$ and in
the R-region the ratio of the coefficients is determined by the boundary
condition (\ref{MITbc}) on the brane:%
\begin{equation}
W_{\nu }(pw)=\left\{
\begin{array}{ll}
J_{\nu }(pw), & 0\leq w\leq w_{0}, \\
G_{\nu _{2},\nu }(pw_{0},pw), & w_{0}\leq w<\infty ,%
\end{array}%
\right.  \label{Wn}
\end{equation}%
with the notation%
\begin{equation}
G_{\mu ,\nu }(x,y)=Y_{\mu }(x)J_{\nu }(y)-J_{\mu }(x)Y_{\nu }(y).  \label{ge}
\end{equation}%
From the boundary condition (\ref{MITbc}) for the modes in the L-region it
follows that the allowed values of the quantum number $p$ are the positive
roots of the equation
\begin{equation}
J_{\nu _{1}}(pw_{0})=0\ .  \label{peqL}
\end{equation}%
The roots with respect to the argument of the Bessel function will be
denoted by $p_{i}=pw_{0}$, $i=1,2,3,\ ...$, assuming that $p_{i+1}>p_{i}$.

The remaining normalization coefficient $C_{\sigma }^{(\pm )}$ in (\ref%
{FermMod}) is determined from the orthonormality condition
\begin{equation}
\int_{0}^{\infty }dr\int_{0}^{\phi _{0}}d\phi \int_{-\infty }^{+\infty
}dz\int dw\,r(a/w)^{4}(\psi _{\sigma }^{(\pm )})^{\dagger }\psi _{\sigma
^{\prime }}^{(\pm )}=\delta _{\sigma \sigma ^{\prime }}\ ,\   \label{nc}
\end{equation}%
where $\delta _{\sigma \sigma ^{\prime }}$ is understood as the Dirac delta
function for continuous quantum numbers in the set $\sigma $ and the
Kronecker delta for discrete ones. The integration over $w$ in (\ref{nc})
goes over $[0,w_{0}]$ in the L-region and over $[w_{0},\infty )$ in the
R-region. We can show that
\begin{equation}
\left\vert C_{\sigma }^{(\pm )}\right\vert ^{2}=\frac{qw_{0}^{-2}\eta
\lambda ^{2}pU_{\nu _{2}}^{\mathrm{(J)}}(pw_{0})}{16\pi ^{2}a^{4}E\sqrt{%
p^{2}+k_{z}^{2}}\kappa _{\eta }b_{\eta }^{(\pm )}}\left\{
\begin{array}{ll}
1, & 0\leq w\leq w_{0}, \\
w_{0}^{2}p, & w_{0}\leq w<\infty ,%
\end{array}%
\right.  \label{Csig}
\end{equation}%
where $\mathrm{J}=\mathrm{L},\mathrm{R}$ specifies the spatial region and%
\begin{eqnarray}
U_{\nu }^{\mathrm{(L)}}(pw_{0}) &=&\frac{2}{J_{\nu }^{2}(pw_{0})},  \notag \\
U_{\nu }^{\mathrm{(R)}}(pw_{0}) &=&\frac{1}{J_{\nu }^{2}(pw_{0})+Y_{\nu
}^{2}(pw_{0})}.  \label{UJ}
\end{eqnarray}%
In the L-region one has $p=p_{i}/w_{0}$.

Here, a comment related to the choice of the coefficients $%
(C_{1},C_{2})=(1,0)$ in the L-region is in order. In the range of the mass $%
ma\geq 1/2$ that choice is dictated by the normalizability condition for the
fermionic modes used in the second quantization procedure. In the range $%
0\leq ma<1/2$, the mode functions with $C_{2}\neq 0$ are normalizable and
for a unique specification of the ratio $C_{2}/C_{1}$ an additional boundary
condition is required on the AdS boundary. Our choice of the modes
corresponds to the setup with two branes located at $w=\varepsilon <w_{0}$
and $w=w_{0}$, where the bag boundary condition is imposed for $%
w=\varepsilon $ and then the limiting transition $\varepsilon \rightarrow 0$
is taken.

\section{Fermionic condensate}

\label{sec:FCond}

\subsection{General formula}

We start our investigation of local VEVs from the FC. It is defined as the
VEV $\langle 0|\bar{\psi}\psi |0\rangle \equiv \langle \bar{\psi}\psi
\rangle $, where $|0\rangle $ stands for the vacuum state. Having the
complete set of the fermionic modes, the FC can be evaluated by the
following mode sum formula:
\begin{equation}
\langle \bar{\psi}\psi \rangle =-\frac{1}{2}\sum_{\sigma }\sum_{\chi
=-,+}\chi \bar{\psi}_{\sigma }^{(\chi )}\psi _{\sigma }^{(\chi )}.
\label{FC}
\end{equation}%
The details of the evaluation procedure for the brane-induced effects do not
depend on the regularization of the divergent expression in the right-hand
side of (\ref{FC}) and we will not specify the corresponding procedure. For
example, we could regularize by the point-splitting technique or by
introducing a cutoff function. Substituting the mode functions (\ref{FermMod}%
) in (\ref{FC}), after the summation over the quantum number $\eta $, we can
see that%
\begin{equation}
\langle \bar{\psi}\psi \rangle _{\mathrm{J}}=-\frac{sqw^{5}}{2\pi
^{2}a^{4}w_{0}^{2}}\int_{0}^{\infty }d\lambda \lambda
\sum_{(p)}\int_{0}^{\infty }dk_{z}\frac{p}{E}\frac{W_{\nu _{1}}(pw)W_{\nu
_{2}}(pw)}{U_{\nu _{2}}^{\mathrm{(J)}}(pw_{0})}\sum_{j}\left[ J_{\beta
_{j}}^{2}(\lambda r)+J_{\beta _{j}+\epsilon _{j}}^{2}(\lambda r)\right] \ ,
\label{FC1}
\end{equation}%
where $\sum_{j}=\sum_{j=\pm 1/2,\pm 3/2,\cdots }$, the functions $W_{\nu
}(pw)$ and $U_{\nu _{2}}^{\mathrm{(J)}}(pw_{0})$ in the L- and R-regions are
defined by (\ref{Wn}) and (\ref{UJ}), $p=p_{i}/w_{0}$ in the L-region and%
\begin{equation}
\sum_{(p)}=\left\{
\begin{array}{ll}
\sum_{i=1}^{\infty }, & 0\leq w\leq w_{0}, \\
w_{0}^{2}\int_{0}^{\infty }dp\,p, & w_{0}\leq w<\infty .%
\end{array}%
\right.  \label{Sump}
\end{equation}%
The parameter $\alpha $, codifying the effects of the magnetic flux, enters
in (\ref{FC1}) through the combination $j+\alpha $. If we present it in the
form $\alpha =N+\alpha _{0}$, with $N$ being an integer and $|\alpha
_{0}|\leq 1/2$, then the integer part is absorbed by the redefinition $%
j+N\rightarrow j$ in the summation over $j$. From here it follows that the
FC\ depends on the fractional part $\alpha _{0}$ only and in (\ref{FC1}) we
can replace $\alpha $ by $\alpha _{0}$.

In order to extract explicitly the contributions induced by the cosmic
staring and by the brane, we substitute in the right-hand side of (\ref{FC1}%
) the representation
\begin{equation}
\frac{1}{E}=\frac{2}{\sqrt{\pi }}\int_{0}^{\infty }dv\ e^{-v^{2}(\lambda
^{2}+p^{2}+k_{z}^{2})}\ .  \label{Erep}
\end{equation}%
The evaluation of the integral over $k_{z}$ is elementary and the $\lambda $%
-integral is evaluated by the formula%
\begin{equation}
\int_{0}^{\infty }d\lambda \lambda e^{-v^{2}\lambda ^{2}}\left[ J_{\beta
_{j}}^{2}(\lambda r)+J_{\beta _{j}+\epsilon _{j}}^{2}(\lambda r)\right] =%
\frac{x}{r^{2}}e^{-x}\left[ I_{\beta _{j}}(x)+I_{\beta _{j}+\epsilon _{j}}(x)%
\right] ,  \label{IntJ1}
\end{equation}%
where $x=r^{2}/(2v^{2})$ and $I_{\nu }(x)$ is the modified Bessel function
\cite{Abra}. Passing to a new integration variable $x$ in the integral over $%
v$, we obtain%
\begin{equation}
\langle \bar{\psi}\psi \rangle _{\mathrm{J}}=-\frac{sqw^{5}}{4\pi
^{2}a^{4}w_{0}^{2}r^{2}}\int_{0}^{\infty }dx\ e^{-x}{\mathcal{J}}(q,\alpha
_{0},x)\sum_{(p)}pe^{-r^{2}p^{2}/(2x)}\frac{W_{\nu _{1}}(pw)W_{\nu _{2}}(pw)%
}{U_{\nu _{2}}^{\mathrm{(J)}}(pw_{0})},  \label{FC2}
\end{equation}%
where
\begin{equation}
{\mathcal{J}}(q,\alpha _{0},x)=\sum_{j}\left[ I_{\beta _{j}}\left( x\right)
+I_{\beta _{j}+\epsilon _{j}}\left( x\right) \right] \ .  \label{Jcal}
\end{equation}

For the further transformation of the expression in the right-hand side of (%
\ref{FC2}) we employ the representation \cite{Beze10}
\begin{eqnarray}
{\mathcal{J}}(q,\alpha _{0},x) &=&\frac{2}{q}e^{x}+\frac{4}{q}%
\sum_{k=1}^{[q/2]}(-1)^{k}c_{k}\cos (2\pi k\alpha _{0})e^{x\cos (2\pi k/q)}
\notag \\
&&+\frac{4}{\pi }\int_{0}^{\infty }du\frac{h(q,\alpha _{0},2u)\sinh u}{\cosh
(2qu)-\cos (q\pi )}e^{-x\cosh 2u},  \label{J-function}
\end{eqnarray}%
where $[q/2]$ represents the integer part of $q/2$, $c_{k}=\cos (\pi k/q)$,
and the function $h(q,\alpha _{0},x)$ is defined as
\begin{equation}
h(q,\alpha _{0},x)=\sum_{l=\pm 1}\cos [\pi q(1/2+l\alpha _{0})]\sinh
[(1/2-l\alpha _{0})qx].  \label{h-function}
\end{equation}%
In the case $1\leq q<2$, the term with the summation on the right-hand side
of (\ref{J-function}) must be omitted. Note that in the absence of the
cosmic string and magnetic flux one has $q=1$, $\alpha _{0}=0$ and ${%
\mathcal{J}}(q,\alpha _{0},x)=2e^{x}$. We will denote the corresponding FC
by $\langle \bar{\psi}\psi \rangle _{\mathrm{J}}^{(0)}$. Now we see that the
part in the FC (\ref{FC2}) coming from the first term in the right-hand side
of (\ref{J-function}) coincides with $\langle \bar{\psi}\psi \rangle _{%
\mathrm{J}}^{(0)}$. In the remaining part the integral over $x$ is expressed
in terms of the modified Bessel function $K_{\nu }(z)$ and one gets%
\begin{eqnarray}
\langle \bar{\psi}\psi \rangle _{\mathrm{J}} &=&\langle \bar{\psi}\psi
\rangle _{\mathrm{J}}^{(0)}-\frac{\pi ^{-2}sw^{5}}{a^{4}w_{0}^{2}r}%
\sum_{(p)}p^{2}\frac{W_{\nu _{1}}(pw)W_{\nu _{2}}(pw)}{U_{\nu _{2}}^{\mathrm{%
(J)}}(pw_{0})}  \notag \\
&&\times \left[ \sum_{k=1}^{[q/2]}(-1)^{k}\frac{c_{k}}{s_{k}}\cos (2\pi
k\alpha _{0})K_{1}(2rps_{k})\right.  \notag \\
&&\left. +\frac{q}{\pi }\int_{0}^{\infty }du\frac{h(q,\alpha _{0},2u)\tanh u%
}{\cosh (2qu)-\cos (q\pi )}K_{1}(2rp\cosh u)\right] ,  \label{FC3}
\end{eqnarray}%
with the notation $s_{k}=\sin (\pi k/q)$.

In order to extract the brane-induced contribution, let us consider the FC
in the brane-free geometry. The corresponding mode functions are given by (%
\ref{FermMod}) with $W_{\nu }(pw)=J_{\nu }(pw)$ and the respective
normalization constant is obtained from (\ref{Csig}) for the region $%
w_{0}\leq w<\infty $ putting $U_{\nu _{2}}^{\mathrm{(J)}}(pw_{0})=1$. The FC
in the brane-free geometry, $\langle \bar{\psi}\psi \rangle _{\mathrm{cs}}^{%
\mathrm{AdS}}$, is obtained from (\ref{FC3}) with the same replacements and
with $\sum_{(p)}$ from (\ref{Sump}) for the R-region. It is decomposed as%
\begin{equation}
\langle \bar{\psi}\psi \rangle _{\mathrm{cs}}^{\mathrm{AdS}}=\langle \bar{%
\psi}\psi \rangle ^{\mathrm{AdS}}+\langle \bar{\psi}\psi \rangle _{\mathrm{cs%
}},  \label{FCfreedec}
\end{equation}%
where $\langle \bar{\psi}\psi \rangle ^{\mathrm{AdS}}$ is the FC\ in pure
AdS spacetime and the part
\begin{eqnarray}
\langle \bar{\psi}\psi \rangle _{\mathrm{cs}} &=&-\frac{sw^{5}}{\pi
^{2}a^{4}r}\int_{0}^{\infty }dp\,p^{3}J_{\nu _{1}}(pw)J_{\nu _{2}}(pw)
\notag \\
&&\times \left[ \sum_{k=1}^{[q/2]}(-1)^{k}\frac{c_{k}}{s_{k}}\cos (2\pi
k\alpha _{0})K_{1}(2rps_{k})\right.  \notag \\
&&\left. +\frac{q}{\pi }\int_{0}^{\infty }du\frac{h(q,\alpha _{0},2u)\tanh u%
}{\cosh (2qu)-\cos (q\pi )}K_{1}(2rp\cosh u)\right] ,  \label{FCfree}
\end{eqnarray}%
is the contribution of the cosmic string in the brane-free geometry (see
\cite{Bell21b}). The integral over $p$ in this expression is expressed in
terms of the associated Legendre function of the second kind and the final
expression for $\langle \bar{\psi}\psi \rangle _{\mathrm{cs}}$ can be found
in \cite{Bell21b}. As seen from (\ref{FCfree}), one has $\langle \bar{\psi}%
\psi \rangle _{\mathrm{cs}}|_{s=-1}=-\langle \bar{\psi}\psi \rangle _{%
\mathrm{cs}}|_{s=+1}$. The FC (\ref{FC3}) is presented as%
\begin{eqnarray}
\langle \bar{\psi}\psi \rangle _{\mathrm{J}} &=&\langle \bar{\psi}\psi
\rangle ^{\mathrm{AdS}}+\langle \bar{\psi}\psi \rangle _{\mathrm{cs}%
}+\langle \bar{\psi}\psi \rangle _{\mathrm{b,J}}^{(0)}  \notag \\
&&-\frac{w^{5}}{\pi ^{2}a^{4}r}\left[ \sum_{k=1}^{[q/2]}(-1)^{k}\frac{c_{k}}{%
s_{k}}\cos (2\pi k\alpha _{0})f_{\mathrm{(J)}}(w_{0},w,2rs_{k})\right.
\notag \\
&&\left. +\frac{q}{\pi }\int_{0}^{\infty }du\frac{h(q,\alpha _{0},2u)\tanh u%
}{\cosh (2qu)-\cos (q\pi )}f_{\mathrm{(J)}}(w_{0},w,2r\cosh u)\right] ,
\label{FC4}
\end{eqnarray}%
where the expressions for the functions $f_{\mathrm{(J)}}(w_{0},w,\gamma )$,
with $\mathrm{J}=\mathrm{R},\mathrm{L}$, are given in Appendix \ref{sec:App1}%
. In (\ref{FC4}), $\langle \bar{\psi}\psi \rangle _{\mathrm{b,J}%
}^{(0)}=\langle \bar{\psi}\psi \rangle _{\mathrm{J}}^{(0)}-\langle \bar{\psi}%
\psi \rangle ^{\mathrm{AdS}}$ is the contribution in the FC induced by the
brane in the geometry where the cosmic string is absent. The latter has been
investigated in \cite{Eliz13} in general number of spatial dimensions $D$
for the case $s=1$. In the special case of (4+1)-dimensional spacetime with $%
D=4$, the corresponding result for $s=\pm 1$ have the form%
\begin{equation}
\langle \bar{\psi}\psi \rangle _{\mathrm{b,J}}^{(0)}=-\frac{w^{5}}{2\pi
^{2}a^{4}}\int_{0}^{\infty }dp\,p^{4}F_{\mathrm{(J)}}(pw_{0},pw),
\label{FCbJ}
\end{equation}%
with the functions%
\begin{eqnarray}
F_{\mathrm{(R)}}(x,y) &=&\frac{I_{\nu _{2}}(x)}{K_{\nu _{2}}(x)}K_{\nu
_{1}}(y)K_{\nu _{2}}(y),  \notag \\
F_{\mathrm{(L)}}(x,y) &=&\frac{K_{\nu _{1}}(x)}{I_{\nu _{1}}(x)}I_{\nu
_{1}}(y)I_{\nu _{2}}(y),  \label{FRL}
\end{eqnarray}%
where $\nu _{1}$ and $\nu _{2}$ are defined by (\ref{n12}).

The further transformation of the functions $f_{\mathrm{(J)}}(w_{0},w,\gamma
)$ in the last term of (\ref{FC4}) is presented in Appendix \ref{sec:App1}.
Substituting (\ref{FCbJ}), (\ref{fR2}) and (\ref{fL2n}) in (\ref{FC4}), the
FC is decomposed as
\begin{equation}
\langle \bar{\psi}\psi \rangle _{\mathrm{J}}=\langle \bar{\psi}\psi \rangle
^{\mathrm{AdS}}+\langle \bar{\psi}\psi \rangle _{\mathrm{cs}}+\langle \bar{%
\psi}\psi \rangle _{\mathrm{b,J}},  \label{FCJdec}
\end{equation}%
where the contribution induced by the brane is given by the formula%
\begin{equation}
\langle \bar{\psi}\psi \rangle _{\mathrm{b,J}}=-\frac{2w^{5}}{\pi ^{2}a^{4}}%
\int_{0}^{\infty }dp\,p^{4}F_{\mathrm{(J)}}(pw_{0},pw)H_{1}(q,\alpha
_{0},2pr).  \label{FCbJ2}
\end{equation}%
In the expression of the right-hand side we have introduced the notation%
\begin{equation}
H_{n}(q,\alpha _{0},x)=\sideset{}{'}{\sum}_{k=0}^{[q/2]}(-1)^{k}c_{k}\cos
(2\pi k\alpha _{0})\frac{J_{n}(xs_{k})}{\left( xs_{k}\right) ^{n}}+\frac{q}{%
\pi }\int_{0}^{\infty }du\frac{h(q,\alpha _{0},2u)\sinh u}{\cosh (2qu)-\cos
(q\pi )}\frac{J_{n}(x\cosh u)}{\left( x\cosh u\right) ^{n}}.  \label{Hnq}
\end{equation}%
Here and below, the prime on the sign of summation means that the term $k=0$
should be taken with additional coefficient 1/2. That term is reduced to $%
2^{-n-1}/\Gamma (n+1)$. Note that in (\ref{FCbJ2}) the contribution of the
term with $k=0$ presents the FC $\langle \bar{\psi}\psi \rangle _{\mathrm{b,J%
}}^{(0)}$, given by (\ref{FCbJ}). For points away from the defect core and
outside of the brane ($r\neq 0$, $w\neq w_{0}$) the renormalization in (\ref%
{FCJdec}) is required for the pure AdS part $\langle \bar{\psi}\psi \rangle
^{\mathrm{AdS}}$ only. In the absence of the defect and brane the background
geometry is maximally symmetric and the renormalized FC $\langle \bar{\psi}%
\psi \rangle ^{\mathrm{AdS}}$ does not depend on the spacetime point.
Comparing the results for the R- and L-regions we see that the formula for
the brane-induced FC in the L-region is obtained from the formula for the
R-region by the replacements $I\rightleftarrows K$, $\nu _{1,2}\rightarrow
\nu _{2,1}$ (see (\ref{FRL})). As seen from (\ref{FCbJ2}), the brane-induced
FC depends on the coordinates $w$, $w_{0}$ and $r$ through the ratios $%
w/w_{0}$ and $r/w$. For the first one we have $w/w_{0}=e^{(y-y_{0})/a}$ and
it determines the physical distance of the observation point from the brane:%
\begin{equation}
|y-y_{0}|=a|\ln (w/w_{0})|.  \label{DistBr}
\end{equation}%
The proper distance from the cosmic string core is given by $r_{p}=ar/w$ and
the ratio $r/w$ presents the proper distance measured in units of the
curvature radius $a$.

The general expression (\ref{FCbJ2}) is further simplified in two special
cases. In the absence of the magnetic flux one has $\alpha _{0}=0$ and
\begin{equation}
h(q,0,2u)=2\cos \left( \pi q/2\right) \sinh \left( qu\right) .
\label{hfalf0}
\end{equation}%
The corresponding expression for the function (\ref{Hnq}) is reduced to
\begin{equation}
H_{n}(q,0,x)=\sideset{}{'}{\sum}_{k=0}^{[q/2]}(-1)^{k}c_{k}\frac{%
J_{n}(xs_{k})}{\left( xs_{k}\right) ^{n}}+\frac{2q}{\pi }\int_{0}^{\infty }du%
\frac{\cos \left( \pi q/2\right) \sinh \left( qu\right) \sinh u}{\cosh
(2qu)-\cos (q\pi )}\frac{J_{n}(x\cosh u)}{\left( x\cosh u\right) ^{n}}.
\label{Hnqalf0}
\end{equation}%
The second special case corresponds to the absence of the planar angle
deficit with $q=1$. In this case
\begin{equation}
h(1,\alpha _{0},2u)=2\sin (\pi \alpha _{0})\sinh \left( 2\alpha _{0}u\right)
\cosh u,  \label{hq1}
\end{equation}%
and the expression for the function (\ref{Hnq}) takes the form%
\begin{equation}
H_{n}(1,\alpha _{0},x)=\frac{2^{-n-1}}{\Gamma (n+1)}+\frac{q}{\pi }\sin (\pi
\alpha _{0})\int_{1}^{\infty }d\tau \,\sinh \left( 2\alpha _{0}\,\mathrm{%
arccosh\,}(\tau )\right) \frac{J_{n}(x\tau )}{x^{n}\tau ^{n+1}}.
\label{Hnq1}
\end{equation}

\begin{figure}[tbph]
\begin{center}
\begin{tabular}{cc}
\epsfig{figure=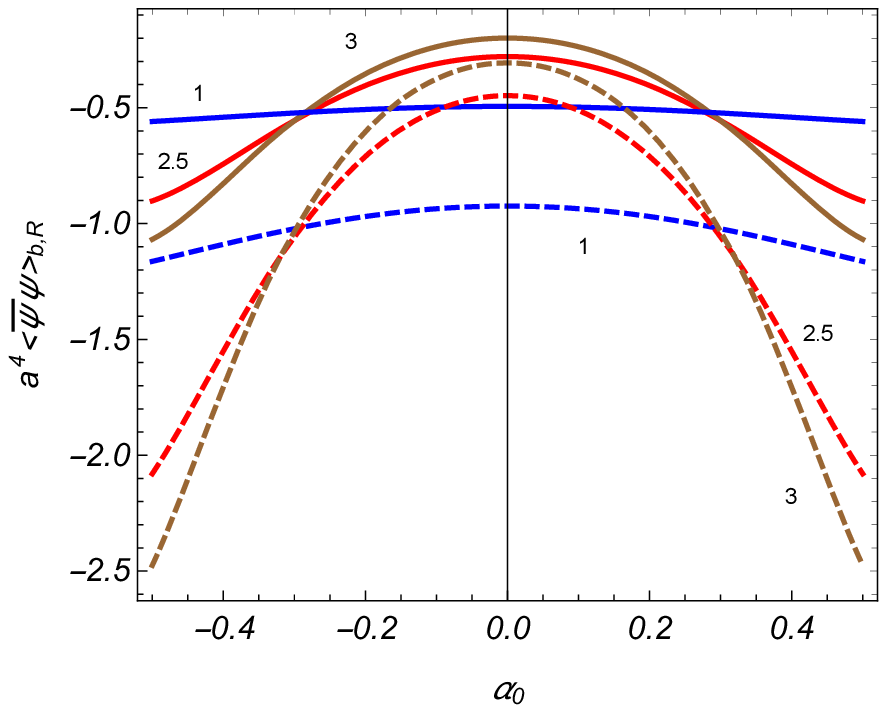,width=7.cm,height=5.5cm} & \quad %
\epsfig{figure=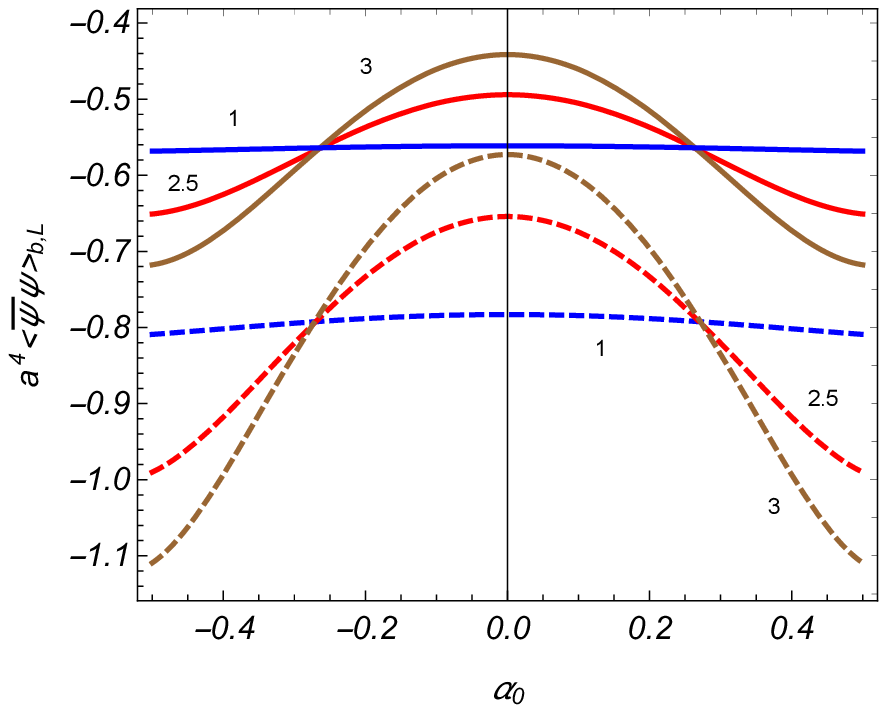,width=7.cm,height=5.5cm}%
\end{tabular}%
\end{center}
\caption{Brane-induced FC versus the fractional part of the ratio of the
magnetic flux to the flux quantum for different values of the parameter $q$
(numbers near the curves). The left and right panels correspond to the R-
and L-regions with $w/w_{0}=1.5$ and $w/w_{0}=0.75$, respectively. The full
and dashed curves present the FC for the fields $s=+1$ and $s=-1$. The
graphs are plotted for $ma=1$ and $r/w_{0}=0.25$.}
\label{figFCalf}
\end{figure}

For general values of the parameters, both the brane-free and brane-induced
contributions in the FC are even periodic functions of the magnetic flux $%
\Phi $ with the period equal to the flux quantum. In Figure \ref{figFCalf}
we have presented the dependence of the brane-induced FC on the fractional
part of the ratio of the magnetic flux to the flux quantum, codified in the
parameter $\alpha _{0}$. The left and right panels correspond to the R- and
L-regions with $w/w_{0}=1.5$ and $w/w_{0}=0.75$, respectively, and the
numbers near the curves are the values of the parameter $q$. The full and
dashed curves present the condensate for the fields with $s=+1$ and $s=-1$.
For the remaining parameters we have taken $ma=1$ and $r/w_{0}=0.25$. As
seen, the dependence on the magnetic flux is stronger for larger values of
the parameter $q$.

\subsection{Special cases, asymptotics and numerical results}

\label{subsec:Spec}

\subsubsection{Minkowski bulk}

First we consider the limit of the Minkowski bulk corresponding to $%
a\rightarrow \infty $ with fixed values of $y$ and $y_{0}$. For the
coordinate $w$ one has $w\approx a+y$. Both the orders and the arguments of
the modified Bessel functions in the expressions for $F_{\mathrm{(J)}%
}(pw_{0},pw)$ are large and we use the corresponding uniform asymptotic
expansions \cite{Abra}. To the leading order this gives%
\begin{equation}
F_{\mathrm{(J)}}(pw_{0},pw)\approx \frac{e^{-2\sqrt{p^{2}+m^{2}}|y-y_{0}|}}{%
2pw\sqrt{p^{2}+m^{2}}}\left( \sqrt{p^{2}+m^{2}}-sm\right) .  \label{FJM}
\end{equation}%
In this case the FC in the R- and L-regions are symmetric. Substituting (\ref%
{FJM}) in (\ref{FCbJ2}) we see that $\lim_{a\rightarrow \infty }\langle \bar{%
\psi}\psi \rangle _{\mathrm{b,J}}=\langle \bar{\psi}\psi \rangle _{\mathrm{b}%
}^{\mathrm{(M)}}$, where the boundary-induced contribution in the Minkowski
bulk is given by%
\begin{equation}
\langle \bar{\psi}\psi \rangle _{\mathrm{b}}^{\mathrm{(M)}}=-\frac{1}{\pi
^{2}}\int_{m}^{\infty }dx\,\left( x^{2}-m^{2}\right) e^{-2x\left(
y-y_{0}\right) }\left( x-sm\right) H_{1}(q,\alpha _{0},2r\sqrt{x^{2}-m^{2}}),
\label{FCbMi}
\end{equation}%
with the function $H_{1}(q,\alpha _{0},x)$ defined in (\ref{Hnq}). The
integral is expressed in terms of the modified Bessel function $K_{\nu }(x)$
(see \cite{Grad}) and
\begin{eqnarray}
\langle \bar{\psi}\psi \rangle _{\mathrm{b}}^{\mathrm{(M)}} &=&-\frac{\sqrt{2%
}m^{4}}{\pi ^{5/2}}\left[ \sideset{}{'}{\sum}_{k=0}^{[q/2]}(-1)^{k}c_{k}\cos
(2\pi k\alpha _{0})F_{\mathrm{(M)}}(2m\left\vert y-y_{0}\right\vert
,2mrs_{k})\right.  \notag \\
&&\left. +\frac{q}{\pi }\int_{0}^{\infty }du\frac{h(q,\alpha _{0},2u)\sinh u%
}{\cosh (2qu)-\cos (q\pi )}F_{\mathrm{(M)}}(2m\left\vert y-y_{0}\right\vert
,2mr\cosh u)\right] .  \label{FCbM}
\end{eqnarray}%
Here, we have introduced the functions%
\begin{equation}
F_{\mathrm{(M)}}(x,u)=xf_{5/2}(\sqrt{x^{2}+u^{2}})-sf_{3/2}(\sqrt{x^{2}+u^{2}%
}),  \label{FM}
\end{equation}%
and%
\begin{equation}
f_{\nu }(x)=x^{-\nu }K_{\nu }(x).  \label{fnu}
\end{equation}%
Note that the function $f_{\nu }(x)$ obeys the relation $f_{\nu }^{\prime
}(x)=-xf_{\nu +1}(x)$. The expression for the FC induced by the cosmic
string in the boundary-free (4+1)-dimensional Minkowski spacetime can be
found in \cite{Bell21b}. For a massless field the latter vanishes. In the
massless limit the boundary-induced contribution (\ref{FCbM}) is reduced to%
\begin{eqnarray}
\langle \bar{\psi}\psi \rangle _{\mathrm{b}}^{\mathrm{(M)}} &=&-\frac{%
3|y-y_{0}|}{16\pi ^{2}}\left\{ \sideset{}{'}{\sum}_{k=0}^{[q/2]}\frac{%
(-1)^{k}c_{k}\cos (2\pi k\alpha _{0})}{\left[ \left( y-y_{0}\right)
^{2}+r^{2}s_{k}^{2}\right] ^{5/2}}\right.  \notag \\
&&\left. +\frac{q}{\pi }\int_{0}^{\infty }du\frac{h(q,\alpha _{0},2u)\sinh u%
}{\cosh (2qu)-\cos (q\pi )}\left[ \left( y-y_{0}\right) ^{2}+r^{2}\cosh ^{2}u%
\right] ^{-5/2}\right\} .  \label{FCbM0}
\end{eqnarray}%
We recall that the $k=0$ term gives the boundary-induced contribution in the
absence of the cosmic string. The divergence of the FC on the boundary comes
from that term alone.

\subsubsection{Massless field}

In the case of a massless field on the AdS bulk the contribution in the FC
induced by a cosmic string in the absence of the boundary is given by \cite%
{Bell21b}%
\begin{eqnarray}
\langle \bar{\psi}\psi \rangle _{\mathrm{cs}} &=&-\frac{3sw^{5}}{16\pi
^{2}a^{4}}\Biggl[\sum_{k=1}^{[q/2]}(-1)^{k}\frac{c_{k}\cos (2\pi k\alpha
_{0})}{(w^{2}+r^{2}s_{k}^{2})^{5/2}}  \notag \\
&&+\frac{q}{\pi }\int_{0}^{\infty }dx\frac{h(q,\alpha _{0},2u)\sinh u}{\cosh
(2qu)-\cos (q\pi )}(w^{2}+r^{2}\cosh ^{2}u)^{-5/2}\Biggr]\ .  \label{FCm0}
\end{eqnarray}%
Comparing with (\ref{FCbM0}), we see that%
\begin{equation}
\langle \bar{\psi}\psi \rangle _{\mathrm{cs}}=(w/a)^{4}\langle \bar{\psi}%
\psi \rangle _{\mathrm{b}}^{\mathrm{(M)}},  \label{FCsm0}
\end{equation}%
where $\langle \bar{\psi}\psi \rangle _{\mathrm{b}}^{\mathrm{(M)}}$ in the
right-hand side is the FC in the locally Minkowski bulk with the line
element (\ref{ds2M}), induced by the boundary at $w=0$. It is given by (\ref%
{FCbM0}) with $y_{0}=0$ and with the replacement $y\rightarrow w$. We could
expect the relation (\ref{FCsm0}), by taking into account that for a
massless fermionic field the boundary-free problem on the AdS bulk is
conformally related to the problem in the Minkowski bulk with a single
boundary at $w=0$. The latter is the conformal image of the AdS boundary.

For a massless field on the AdS bulk one has $\nu _{1}=-s/2$, $\nu _{2}=s/2$
and the modified Bessel functions in (\ref{FRL}) are expressed in terms of
the elementary functions. This gives:%
\begin{eqnarray}
F_{\mathrm{(R)}}(pw_{0},pw) &=&\frac{e^{2pw_{0}}-s}{2pwe^{2pw}},  \notag \\
F_{\mathrm{(L)}}(pw_{0},pw) &=&\frac{\sinh \left( 2pw\right) }{%
pw(e^{2pw_{0}}+s)}.  \label{FRLm0}
\end{eqnarray}%
In the expression (\ref{FCbJ2}) for the R-region the integral is evaluated
by using the formula from \cite{Prud2}:%
\begin{equation}
\int_{0}^{\infty }dp\,p^{3}F_{\mathrm{(R)}}(pw_{0},pw)J_{1}(2xp)=\frac{3x}{16%
}\left\{ \frac{1-w_{0}/w}{[\left( w-w_{0}\right) ^{2}+x^{2}]^{5/2}}-\frac{s}{%
\left( w^{2}+x^{2}\right) ^{5/2}}\right\} .  \label{IntR}
\end{equation}%
The contribution to the brane-induced part $\langle \bar{\psi}\psi \rangle _{%
\mathrm{b,R}}$ coming from the second term in the figure brackets of (\ref%
{IntR}) with $k\neq 0$ is cancelled by the brane-free term $\langle \bar{\psi%
}\psi \rangle _{\mathrm{cs}}$, given by (\ref{FCm0}), and the total
condensate in the R-region is presented as
\begin{eqnarray}
\langle \bar{\psi}\psi \rangle _{\mathrm{R}} &=&\langle \bar{\psi}\psi
\rangle ^{\mathrm{AdS}}+\frac{3s}{32\pi ^{2}a^{4}}-\frac{3w^{4}\left(
w-w_{0}\right) }{16\pi ^{2}a^{4}}\left\{ \sideset{}{'}{\sum}_{k=0}^{[q/2]}%
\frac{(-1)^{k}c_{k}\cos (2\pi k\alpha _{0})}{[\left( w-w_{0}\right)
^{2}+r^{2}s_{k}^{2}]^{5/2}}\right.  \notag \\
&&\left. +\frac{q}{\pi }\int_{0}^{\infty }du\frac{h(q,\alpha _{0},2u)\sinh u%
}{\cosh (2qu)-\cos (q\pi )}[\left( w-w_{0}\right) ^{2}+r^{2}\cosh
^{2}u]^{-5/2}\right\} .  \label{FCRm0}
\end{eqnarray}%
Comparing with (\ref{FCbM0}) (with the replacements $y,y_{0}\rightarrow
w,w_{0}$) we see that the last term in (\ref{FCRm0}) is conformally related
to the FC in the Minkowski bulk (\ref{ds2M}) induced by the boundary at $%
w=w_{0}$.

In the massless case, the expression for the FC in the L-region is obtained
substituting (\ref{FRLm0}) in (\ref{FCbJ2}). Compared to the R-region, the
corresponding expression is more complicated. That is related to the fact
that the probem in the L-region is conformally related to the problem in the
Minkowski with two planar boundaries located at $w=0$ and $w=w_{0}$. The
former is the conformal image of the AdS boundary.

\subsubsection{Asymptotics with respect to the distance from the brane}

First let us consider the FC in the R-region at large distances from the
brane, $w\gg w_{0}$, that corresponds to $(y-y_{0})\gg a$. The dominant
contribution to the integral over $p$ in (\ref{FCbJ2}) comes from the region
where $pw_{0}\ll 1$. Expanding the corresponding modified Bessel functions
one gets%
\begin{equation}
\langle \bar{\psi}\psi \rangle _{\mathrm{b,R}}\approx -\frac{%
4(w_{0}/2w)^{\nu _{2}+|\nu _{2}|}}{\pi ^{2}a^{4}\Gamma (|\nu _{2}|)\Gamma
\left( 1+\nu _{2}\right) }\int_{0}^{\infty }dp\,p^{4+\nu _{2}+|\nu
_{2}|}K_{\nu _{1}}(p)K_{\nu _{2}}(p)H_{1}(q,\alpha _{0},2pr/w).
\label{FCRwlarge}
\end{equation}%
If, in addition, $w\gg r$, further simplification is made by expanding the
integrand with respect to $r/w$ (for the near string asymptotic see below).
From (\ref{FCRwlarge}) we see that the leading term in the brane-induced FC $%
\langle \bar{\psi}\psi \rangle _{\mathrm{b,R}}$ behaves as $(w_{0}/w)^{\nu
_{2}+|\nu _{2}|}$. This shows that for $s=1$ the brane-induced FC decays as $%
(w_{0}/w)^{2ma+1}$. For the field with $s=-1$, depending on the mass, two
qualitatively different cases are realized. The first one corresponds to the
range of the mass $ma>1/2$ and in this case the condensate $\langle \bar{\psi%
}\psi \rangle _{\mathrm{b,R}}$ tends to zero like $(w_{0}/w)^{2ma-1}$. For
the masses in the range $ma<1/2$ and for fixed $r/w$ the brane-induced
contribution tends to a finite nonzero value in the limit $%
w_{0}/w\rightarrow 0$. Note that the asymptotic (\ref{FCRwlarge}) also
determines the behavior of the FC when the location of the brane tends to
the AdS boundary, $w_{0}\rightarrow 0$, for fixed values of $w$ and $r$.

In the L-region, at large distances from the brane one has $w\ll w_{0}$. For
fixed values of $w_{0}$ and $r$ this corresponds to points near the AdS
boundary. In this limit, in the expression (\ref{FCbJ2}) for the L-region we
introduce a new integration variable $u=pw_{0}$ and expand the modified
Bessel functions with the arguments $uw/w_{0}$. Keeping the leading order
terms we get%
\begin{equation}
\langle \bar{\psi}\psi \rangle _{\mathrm{b,L}}\approx -\frac{\left(
3+2ma\right) (w/w_{0})^{5+2ma}}{2^{2ma}\pi ^{2}a^{4}\Gamma ^{2}\left(
3/2+ma\right) }\int_{0}^{\infty }dp\,p^{4+2ma}\frac{K_{\nu _{1}}(p)}{I_{\nu
_{1}}(p)}H_{1}(q,\alpha _{0},2pr/w_{0}),  \label{FCLnearBound}
\end{equation}%
and the brane-induced FC\ vanishes on the AdS boundary like $\left(
w/w_{0}\right) ^{2ma+5}$. Note that near the AdS boundary the brane-free
contribution behaves as \cite{Bell21b}
\begin{equation}
\langle \bar{\psi}\psi \rangle _{\mathrm{cs}}\approx -\frac{\left(
1/2+ma\right) (3/2+ma)}{2^{2+2ma}\pi ^{2}a^{4}(r/w)^{5+2ma}}%
h_{5+2ma}(q,\alpha _{0})\ ,  \label{larger}
\end{equation}%
with the notation%
\begin{equation}
h_{n}(q,\alpha _{0})=\sum_{k=1}^{[q/2]}\frac{(-1)^{k}c_{k}}{s_{k}^{n}}\cos
(2\pi k\alpha _{0})+\frac{q}{\pi }\int_{0}^{\infty }dx\,\frac{h(q,\alpha
_{0},2x)\sinh (x)\cosh ^{-n}x}{\cosh (2qx)-\cos (q\pi )}\ .  \label{hn}
\end{equation}%
Hence, in the limit $w\rightarrow 0$ both the contributions tend to zero
like $w^{5+2ma}$.

Now we turn to the near brane asymptotic. The brane-induced FC\ diverges for
points on the brane, $w=w_{0}$. As it is seen from the representation (\ref%
{FC3}), the divergence comes from the $k=0$ term that corresponds to the FC
in the geometry where the cosmic string is absent. The remaining
contributions are finite on the brane due to the presence of the modified
Bessel function $K_{1}(x)$. Moreover, from (\ref{FC3}) it follows that for $%
r>0$ the difference $\langle \bar{\psi}\psi \rangle _{\mathrm{J}}-\langle
\bar{\psi}\psi \rangle _{\mathrm{J}}^{(0)}$ vanishes on the brane. Note that
in the representation (\ref{FCJdec}), with the brane-induced contribution
from (\ref{FCbJ2}), for the evaluation of that difference on the brane the
direct substitution $w=w_{0}$ in the corresponding integrand is not allowed.
For points near the brane the dominant contribution to the integral for the $%
k=0$ term in (\ref{FCbJ2}) comes from the integration range where the
arguments of the modified Bessel functions are large. By making use of the
corresponding asymptotic expressions \cite{Abra}, we can see that to the
leading order $F_{\mathrm{(J)}}(pw_{0},pw)\approx e^{-2p|w-w_{0}|}/(2pw)$.
After evaluating the integral we get the leading behavior:%
\begin{equation}
\langle \bar{\psi}\psi \rangle _{\mathrm{b,J}}\approx -\frac{3\left(
1-w_{0}/w\right) ^{-4}}{32\pi ^{2}a^{4}}.  \label{FCnearBr}
\end{equation}%
Note that for points near the brane one has $|y-y_{0}|\ll a$ and $%
1-w_{0}/w\approx (y-y_{0})/a$. The leading term is written as $\langle \bar{%
\psi}\psi \rangle _{\mathrm{b,J}}\approx 3\left( y-y_{0}\right) ^{-4}/(32\pi
^{2})$. It coincides with the corresponding result in the Minkowski bulk and
the effects of the spacetime curvature are weak. That is related to the fact
that near the brane the main contribution to the VEVs comes from
fluctuations with small wavelengths and the influence of gravity on those
fluctuations is weak. The leading term (\ref{FCnearBr}) does not depend on
the planar angle deficit and in the region near the brane the effects of the
cosmic string are subdominant.

\begin{figure}[tbph]
\begin{center}
\begin{tabular}{cc}
\epsfig{figure=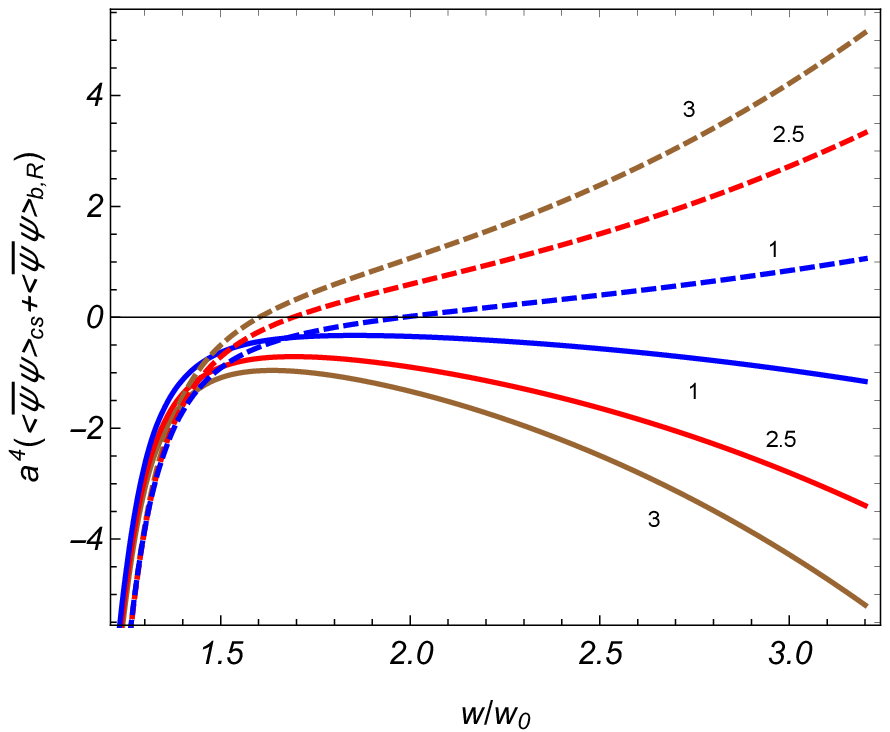,width=7.cm,height=5.5cm} & \quad %
\epsfig{figure=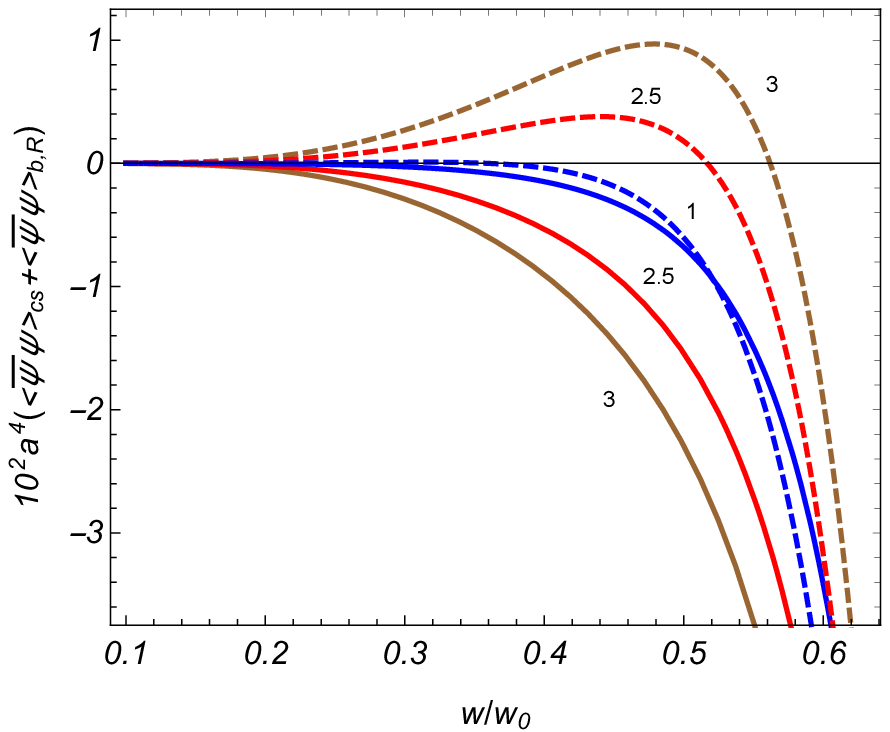,width=7.cm,height=5.5cm}%
\end{tabular}%
\end{center}
\caption{The FC induced by the cosmic string and the brane versus the ratio $%
w/w_{0}$ for different values of the parameter $q$ (numbers near the
curves). The full and dashed curves correspond to $s=+1$ and $s=-1$,
respectively. For the other parameters we have taken $ma=1$, $r/w=0.25$ and $%
\protect\alpha _{0}=0.3$. }
\label{figFCw}
\end{figure}

Figure \ref{figFCw} presents the dependence of the FC, induced by the cosmic
string and brane, as a function of the ratio $w/w_{0}$ that determines the
distance from the brane (see (\ref{DistBr})). The left and right panels
correspond to the R- and L-regions and the full and dashed curves correspond
to the fields with $s=+1$ and $s=-1$. The graphs are plotted for $ma=1$, $%
r/w=0.25$, $\alpha _{0}=0.3$ and the numbers near the curves are the values
of the parameter $q$. The graphs on Figure \ref{figFCw} confirm the features
clarified by the asymptotic analysis: near the brane the FC is dominated by
the brane-induced contribution and the dependence on $q$ is weak, whereas at
large distances and near the horizon the brane-free contribution dominates.
As seen, depending on the distance from the brane, the FC may change the
sign.

\subsubsection{Small and large distances from the cosmic string}

We finish the qualitative description of the behavior for the FC considering
small and large distance asymptotics with respect to the cosmic string core.
At small proper distances $r_{p}$ from the cosmic string core, compared with
the curvature radius, one has $r/w\ll 1$, and the brane-free contribution
for a massive field behaves as \cite{Bell21b}%
\begin{equation}
\langle \bar{\psi}\psi \rangle _{\mathrm{cs}}\approx -\frac{mh_{3}(q,\alpha
_{0})}{8\pi ^{2}(ar/w)^{3}}\ .  \label{FConcs}
\end{equation}%
It diverges on the cosmic string as the inverse cube of the distance. For a
massless field the condensate $\langle \bar{\psi}\psi \rangle _{\mathrm{cs}}$
is finite on the string for $2|\alpha _{0}|<1-1/q$ with the value $\langle
\bar{\psi}\psi \rangle _{\mathrm{cs}}|_{r=0}=-3h_{0}(q,\alpha _{0})/(16\pi
^{2}a^{4})$. For $2|\alpha _{0}|>1-1/q$ the brane-free part $\langle \bar{%
\psi}\psi \rangle _{\mathrm{cs}}$ for a massless field diverges on the
string like $\left( w/r\right) ^{1-\left( 1-2|\alpha _{0}|\right) q}$. For $%
2|\alpha _{0}|<1-1/q$ the limiting value for the brane-induced contribution $%
\langle \bar{\psi}\psi \rangle _{\mathrm{b,J}}$ on the string is obtained
directly from (\ref{FCbJ2}) putting in the integrand $r=0$:%
\begin{equation}
\langle \bar{\psi}\psi \rangle _{\mathrm{b,J}}|_{r=0}=-\frac{%
1+2h_{0}(q,\alpha _{0})}{2\pi ^{2}a^{4}}\int_{0}^{\infty }dx\,x^{4}F_{%
\mathrm{(J)}}(xw_{0}/w,x).  \label{FCbonst}
\end{equation}%
Comparing with the FC in the geometry where the cosmic string is absent,
given by (\ref{FCbJ}), the following relation is seen: $\langle \bar{\psi}%
\psi \rangle _{\mathrm{b,J}}|_{r=0}=[1+2h_{0}(q,\alpha _{0})]\langle \bar{%
\psi}\psi \rangle _{\mathrm{b,J}}^{(0)}$. From (\ref{J-function}) for the
factor in this expression one has%
\begin{equation}
1+2h_{0}(q,\alpha _{0})=\frac{q}{2}\lim_{x\rightarrow 0}e^{-x}{\mathcal{J}}%
(q,\alpha _{0},x).  \label{h0}
\end{equation}%
Now, by taking into account (\ref{Jcal}), we see that for $2|\alpha
_{0}|<1-1/q$ the right-hand side of (\ref{h0}) tends to zero as $x^{q\left(
1/2-|\alpha _{0}|\right) -1/2}$. Hence, we conclude that $\langle \bar{\psi}%
\psi \rangle _{\mathrm{b,J}}|_{r=0}=0$ for $2|\alpha _{0}|<1-1/q$. In the
range $2|\alpha _{0}|>1-1/q$ the contribution $\langle \bar{\psi}\psi
\rangle _{\mathrm{b,J}}$ diverges on the string. The divergence comes from
the integral over $u$ in (\ref{FCbJ2}). For points close to the string the
dominant contribution to that integral comes from large values of $u$. By
using the corresponding asymptotic, after evaluating the integral, to the
leading order we get%
\begin{equation}
\langle \bar{\psi}\psi \rangle _{\mathrm{b,J}}\approx -\frac{qa^{-4}\left(
r/2w\right) ^{2\alpha _{q}-1}}{2\pi ^{2}(1+2\alpha _{q})\Gamma
^{2}(1/2+\alpha _{q})}\int_{0}^{\infty }dp\,p^{3+2\alpha _{q}}F_{\mathrm{(J)}%
}(pw_{0}/w,p),  \label{FCnearSt}
\end{equation}%
where $\alpha _{q}=(1/2-|\alpha _{0}|)q$. Note that under the condition $%
2|\alpha _{0}|>1-1/q$ one has $\alpha _{q}<1/2$ and the condensate (\ref%
{FCnearSt}) is negative near the string. From (\ref{h0}) it follows that $%
1+2h_{0}(q,\alpha _{0})=q/2$ for $\alpha _{q}=1/2$. The limiting value of
the FC on the cosmic string for this special case is obtained from (\ref%
{FCbonst}) with that replacement. The same result is also obtained from (\ref%
{FCnearSt}) in the limit $\alpha _{q}\rightarrow 1/2$.

In the discussion of the large distance asymptotic it is more convenient to
use the representation (\ref{FC3}), where the last term is induced by the
presence of the cosmic string. The behavior of the FC in that region is
qualitatively different for the L- and R-regions and we consider them
separately. In the L-region the spectrum of the quantum number $p$ is
discrete and at large distances the main contribution to the string-induced
part comes from the term with the lowest mode $p=p_{1}/w_{0}$. For $q>2$ the
term $k=1$ dominates and the cosmic string induced effects, given by $%
\langle \bar{\psi}\psi \rangle _{\mathrm{L}}-\langle \bar{\psi}\psi \rangle
_{\mathrm{L}}^{(0)}$, are suppressed by the factor $e^{-2p_{1}s_{1}r/w_{0}}$%
. For $1\leq q<2$ the sum over $k$ is absent in (\ref{FC3}) and we need to
estimate the integral. In the region under consideration the integral is
dominated by the integration range near the lower limit. By expanding the
integrand for small values of $u$, we can see that $\langle \bar{\psi}\psi
\rangle _{\mathrm{L}}-\langle \bar{\psi}\psi \rangle _{\mathrm{L}%
}^{(0)}\propto e^{-2\left( r/w_{0}\right) p_{1}}$ and in this case the
suppression is stronger.

In the R-region the spectrum of the quantum number is continuous and $%
\sum_{(p)}$ in (\ref{FC3}) is defined in accordance with (\ref{Sump}). At
large distances from the cosmic string the dominant contribution to the
corresponding integral over $p$ comes from the region near the lower limit
of the integration and we can use the asymptotic formulas for the Bessel and
Neumann functions for small arguments \cite{Abra}. By using the integral
\cite{Prud2}
\begin{equation}
\int_{0}^{\infty }dx\,x^{\alpha -1}K_{\nu }(cx)=\frac{2^{\alpha -2}}{%
c^{\alpha }}\Gamma \left( \frac{\alpha +\nu }{2}\right) \Gamma \left( \frac{%
\alpha -\nu }{2}\right) ,  \label{IntK}
\end{equation}%
we can see that%
\begin{equation}
\langle \bar{\psi}\psi \rangle _{\mathrm{R}}\approx \langle \bar{\psi}\psi
\rangle _{\mathrm{R}}^{(0)}-\frac{2s\left[ 1-(w_{0}/w)^{1+2sma}\right] }{\pi
^{2}a^{4}(2r/w)^{5+2sma}}\left( 3+2sma\right) \left( 1+2sma\right)
h_{5+2sma}(q,\alpha _{0}),  \label{FCRlarg1}
\end{equation}%
for the cases $s=1$ and $s=-1$, $ma<1/2$. In the case $s=-1$, $ma>1/2$ the
asymptotic behavior is described by the expression%
\begin{equation}
\langle \bar{\psi}\psi \rangle _{\mathrm{R}}\approx \langle \bar{\psi}\psi
\rangle _{\mathrm{R}}^{(0)}+\frac{2\left( 4m^{2}a^{2}-1\right) \left[
1-(w_{0}/w)^{2ma-1}\right] }{\pi ^{2}a^{4}\left( w/w_{0}\right)
^{2ma-1}(2r/w)^{3+2ma}}h_{3+2ma}(q,\alpha _{0}).  \label{FClarg2}
\end{equation}%
These estimates show that, unlike the L-region, the decay of the topological
contributions in the R-region, as functions of the proper distance from the
cosmic string, exhibit an inverse power-law decrease at large distances.

\begin{figure}[tbph]
\begin{center}
\begin{tabular}{cc}
\epsfig{figure=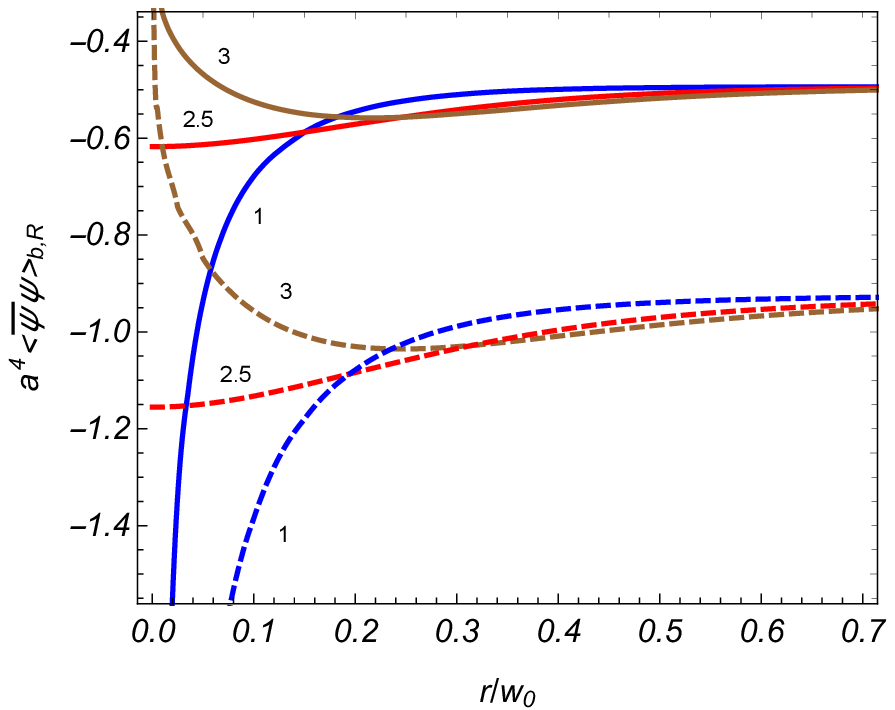,width=7.cm,height=5.5cm} & \quad %
\epsfig{figure=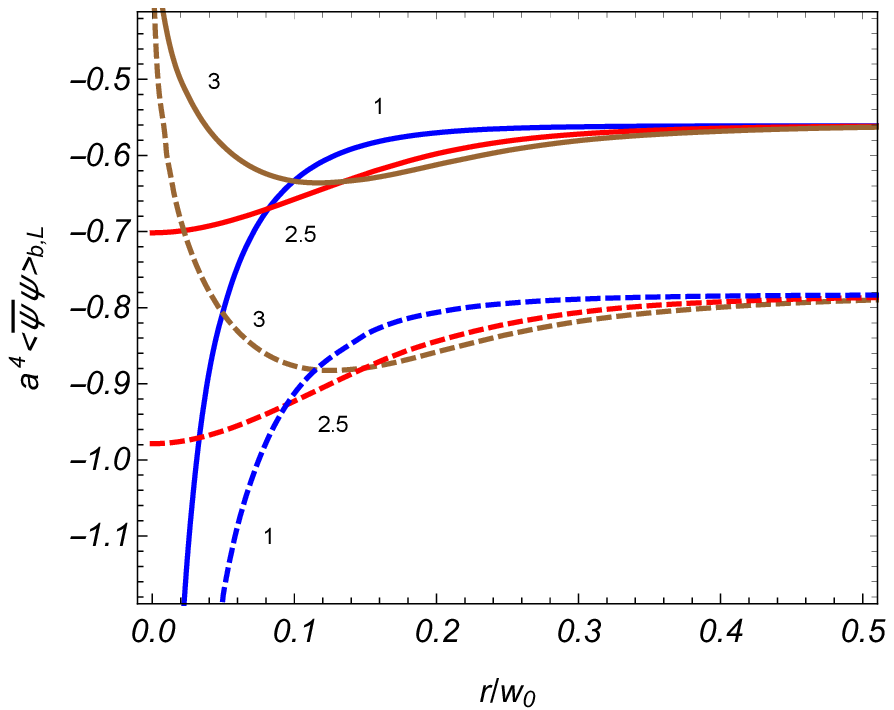,width=7.cm,height=5.5cm}%
\end{tabular}%
\end{center}
\caption{Brane-induced FC as a function of the radial distance from the
cosmic string for different values of the parameter $q$ (numbers near the
curves). The left and right panels correspond to the R- and L-regions with $%
w/w_{0}=1.5$ and $w/w_{0}=0.75$, respectively. The full and dashed curves
present the FC for the fields $s=+1$ and $s=-1$. The graphs are plotted for $%
ma=1$ and $\protect\alpha _{0}=0.3$.}
\label{figFCr}
\end{figure}

In Figure \ref{figFCr} we have displayed the brane-induced contribution to
the FC as a function of the ratio $r/w_{0}$ for the R- and L-regions (left
and right panels, respectively). The full and dashed curves correspond to
the fields with $s=1$ and $s=-1$. The graphs are plotted for $ma=1$, $\alpha
_{0}=0.3$ and the numbers near the curves correspond to the values of the
parameter $q$. The graphs on the left and right panels are plotted for $%
w/w_{0}=1.5$ and $w/w_{0}=0.75$, respectively. Note that the case $q=2.5$
corresponds to the critical value $\alpha _{q}=1/2$ for which the
brane-induced FC takes a finite nonzero value on the cosmic string. In
accordance with (\ref{FCnearSt}), for $q=1$ the condensate $\langle \bar{\psi%
}\psi \rangle _{\mathrm{b,J}}$ diverges on the string. For $q=3$ the
brane-induced condensate tends to zero in the limit $r\rightarrow 0$. At
large distances from the cosmic string one has $\langle \bar{\psi}\psi
\rangle _{\mathrm{J}}\rightarrow \langle \bar{\psi}\psi \rangle _{\mathrm{J}%
}^{(0)}$.

The results obtained above can be used to estimate the effects of the
nonzero FC in models with self-interacting fermions or with fermions
interacting with other quantum fields. Examples are the fermionic fields
with Nambu-Jona-Lasinio type self-interaction (described by the Lagrangian
density proportional to $\left( \bar{\psi}\psi \right) ^{2}$) and the
fermions interacting with a scalar fields $\varphi $ through the Lagrangian
density proportional to $\bar{\psi}\psi \varphi ^{2}$. The formation of the
FC gives rise to mass terms in the field equations proportional to $\langle
\bar{\psi}\psi \rangle \psi $ and $\langle \bar{\psi}\psi \rangle \varphi $,
for the fermionic and scalar fields, respectively. To the leading order with
respect to the interactions, the FC in those terms is given by the quantity
evaluated within the framework of the free field theory. For the geometry
under consideration that quantity is determined by the expressions given in
this section. Note that, depending on the sign of the FC, the generation of
additional mass terms may induce instabilities in the respective field
theories.

\section{Vacuum expectation value of the energy-momentum tensor}

\label{sec:EMT0}

In this section we investigate another important characteristic of the
fermionic vacuum, the VEV of the energy-momentum tensor $\left\langle
0|T_{\mu \nu }|0\right\rangle \equiv \left\langle T_{\mu \nu }\right\rangle $%
. It is evaluated by using the mode sum formula
\begin{equation}
\left\langle T_{\mu \nu }\right\rangle =-\frac{i}{4}\sum_{\sigma }\sum_{\chi
=-,+}\chi {\left[ \bar{\psi}_{\sigma }^{(\chi )}\gamma _{(\mu }\mathcal{D}%
_{\nu )}\psi _{\sigma }^{(\chi )}-(\mathcal{D}_{(\mu }\bar{\psi}_{\sigma
}^{(\chi )})\gamma _{\nu )}\psi _{\sigma }^{(\chi )}\right] }\ .  \label{EMT}
\end{equation}%
where the covariant derivative operator acting on the Dirac adjoint spinor
is given by $\mathcal{D}_{\mu }{\bar{\psi}}=\partial _{\mu }{\bar{\psi}}%
-ieA_{\mu }{\bar{\psi}}-{\bar{\psi}}\Gamma _{\mu }$ and the brackets in the
index expression mean the symmetrization over the enclosed indices.

\subsection{General expressions}

Inserting the mode functions (\ref{FermMod}), using the relation $\{\gamma
_{\mu },\Gamma _{\mu }\}=0$, and summing over $\chi $ and $\eta $, we can
see that the VEVs for the off-diagonal components vanish. The VEVs for the
diagonal components are presented in the combined form (no summation over $%
\mu $)%
\begin{equation}
\left\langle T_{\mu }^{\mu }\right\rangle _{\mathrm{J}}=-\frac{qw^{6}}{4\pi
^{2}a^{5}w_{0}^{2}}\int_{0}^{\infty }d\lambda \lambda
\sum_{(p)}\int_{0}^{\infty }dk_{z}\,\left( k_{z}^{2}\right) ^{\delta _{4\mu
}}\frac{E^{2\delta _{0\mu }}}{E}\frac{W_{\nu _{1}}^{(\mu )}(pw)}{U_{\nu
_{2}}^{\mathrm{(J)}}(pw_{0})}\sum_{j}R_{\beta _{j}}^{(\mu )}\left( \lambda
r\right) ,  \label{TmuJ}
\end{equation}%
where, as before, $\mathrm{J}=\mathrm{R},\mathrm{L}$ for the R- and
L-regions, respectively. The functions for the separate components are
defined by the expressions%
\begin{eqnarray}
R_{\beta _{j}}^{(0)}\left( \lambda r\right) &=&R_{\beta _{j}}^{(3)}\left(
\lambda r\right) =-R_{\beta _{j}}^{(4)}\left( \lambda r\right) =J_{\beta
_{j}}^{2}(\lambda r)+J_{\beta _{j}+\epsilon _{j}}^{2}(\lambda r),  \notag \\
R_{\beta _{j}}^{(1)}\left( \lambda r\right) &=&\epsilon _{j}\lambda ^{2}
\left[ J_{\beta _{j}}^{\prime }(\lambda r)J_{\beta _{j}+\epsilon
_{j}}(\lambda r)-J_{\beta _{j}}(\lambda r)J_{\beta _{j}+\epsilon
_{j}}^{\prime }(\lambda r)\right] ,  \notag \\
R_{\beta _{j}}^{(2)}\left( \lambda r\right) &=&-\frac{\lambda }{r}(2\beta
_{j}+\epsilon _{j})J_{\beta _{j}}(\lambda r)J_{\beta _{j}+\epsilon
_{j}}(\lambda r),  \label{R2}
\end{eqnarray}%
and%
\begin{eqnarray}
W_{\nu _{1}}^{(\mu )}(pw) &=&W_{\nu _{1}}^{2}(pw)+W_{\nu _{2}}^{2}(pw),\;\mu
=0,1,2,4,  \notag \\
W_{\nu _{1}}^{(3)}(pw) &=&sp^{2}\left[ W_{\nu _{1}}^{\prime }(pw)W_{\nu
_{2}}(pw)-W_{\nu _{1}}(pw)W_{\nu _{2}}^{\prime }(pw)\right] .  \label{W3}
\end{eqnarray}%
The remaining notations are the same as those in (\ref{FC1}). Note that we
have the relations%
\begin{eqnarray}
R_{\beta _{j}}^{(1)}\left( \lambda r\right) &=&-\lambda ^{2}R_{\beta
_{j}}^{(0)}\left( \lambda r\right) -R_{\beta _{j}}^{(2)}\left( \lambda
r\right) ,  \notag \\
W_{\nu _{1}}^{(3)}(pw) &=&-p^{2}W_{\nu _{1}}^{(0)}(pw)+2ma\frac{p}{w}W_{\nu
_{1}}(pw)W_{\nu _{2}}(pw).  \label{R1}
\end{eqnarray}%
It can be checked that the components (\ref{TmuJ}) obey the trace relation
\begin{equation}
\left\langle T_{\mu }^{\mu }\right\rangle _{\mathrm{J}}=sm\langle \bar{\psi}%
\psi \rangle _{\mathrm{J}}.  \label{Trace}
\end{equation}

The further transformations of (\ref{TmuJ}) are similar to those for the FC.
By using the integral representation (\ref{Erep}) and integrating over $%
k_{z} $ we get (no summation over $\mu $)%
\begin{equation}
\left\langle T_{\mu }^{\mu }\right\rangle _{\mathrm{J}}=-\frac{qw^{6}}{4\pi
^{2}a^{5}w_{0}^{2}}\sum_{j}\int_{0}^{\infty }dv\ \left( -\partial
_{v^{2}}\right) ^{\delta _{0\mu }}\sum_{(p)}\,\frac{e^{-v^{2}p^{2}}}{%
v(2v^{2})^{\delta _{4\mu }}}\frac{W_{\nu _{1}}^{(\mu )}(pw)}{U_{\nu _{2}}^{%
\mathrm{(J)}}(pw_{0})}\int_{0}^{\infty }d\lambda \lambda e^{-v^{2}\lambda
^{2}}R_{\beta _{j}}^{(\mu )}\left( \lambda r\right) .  \label{TmuJ1}
\end{equation}%
For the components with $\mu =0,3,4$ the integrals over $\lambda $ are
evaluated by the formula (\ref{IntJ1}). For the component $\mu =2$ we use
the formula%
\begin{equation}
\int_{0}^{\infty }d\lambda \,\lambda ^{2}e^{-\lambda ^{2}v^{2}}J_{\beta
_{j}}(\lambda r)J_{\beta _{j}+\epsilon _{j}}(\lambda r)=\frac{\epsilon _{j}}{%
r^{3}}x^{2}e^{-x}\left[ I_{\beta _{j}}(x)-I_{\beta _{j}+\epsilon _{j}}(x)%
\right] \ ,  \label{IntJ2}
\end{equation}%
and the relation
\begin{equation}
I_{\beta _{j}}(x)-I_{\beta _{j}+\epsilon _{j}}(x)=\frac{x\partial _{x}-x+1/2%
}{\epsilon _{j}\beta _{j}+1/2}\left[ I_{\beta _{j}}(x)+I_{\beta
_{j}+\epsilon _{j}}(x)\right] \ ,  \label{Irel}
\end{equation}%
with $x=r^{2}/(2v^{2})$. The remaining integral over $\lambda $ for $\mu =1$
is obtained by using the relation (\ref{R1}). After integrating by parts the
energy density, the VEV is presented as (no summation over $\mu $)%
\begin{equation}
\left\langle T_{\mu }^{\mu }\right\rangle _{\mathrm{J}}=\frac{(-1)^{\delta
_{3\mu }}qw^{6}}{8\pi ^{2}a^{5}w_{0}^{2}r^{4-2\delta _{3\mu }}}%
\int_{0}^{\infty }dx\ x^{1-\delta _{3\mu }}\sum_{(p)}\,e^{-r^{2}p^{2}/2x}%
\frac{W_{\nu _{1}}^{(\mu )}(pw)}{U_{\nu _{2}}^{\mathrm{(J)}}(pw_{0})}\left(
2x\partial _{x}+1\right) ^{\delta _{2\mu }}e^{-x}{\mathcal{J}}(q,\alpha
_{0},x),  \label{TmuJ2}
\end{equation}%
with the notation (\ref{J-function}). From here it follows that
\begin{equation}
\left\langle T_{0}^{0}\right\rangle _{\mathrm{J}}=\left\langle
T_{1}^{1}\right\rangle _{\mathrm{J}}=\left\langle T_{4}^{4}\right\rangle _{%
\mathrm{J}},  \label{T014}
\end{equation}%
and%
\begin{equation}
\left\langle T_{2}^{2}\right\rangle _{\mathrm{J}}=(r\partial
_{r}+1)\left\langle T_{0}^{0}\right\rangle _{\mathrm{J}}.  \label{T2}
\end{equation}%
Based on this relations, we continue our discussion for the components $\mu
=0,3$.

For the factor ${\mathcal{J}}(q,\alpha _{0},x)$ in the integrand of (\ref%
{TmuJ2}) we have the representation (\ref{J-function}). The part coming from
the first term in the right-hand side gives the VEV in the geometry without
a cosmic string, denoted here by $\left\langle T_{\mu }^{\mu }\right\rangle
_{\mathrm{J}}^{(0)}$. In the part induced by the presence of the string the
integral over $x$ is expressed in terms of the modified Bessel function and
we get (no summation over $\mu =0,3$)
\begin{eqnarray}
\left\langle T_{\mu }^{\mu }\right\rangle _{\mathrm{J}} &=&\left\langle
T_{\mu }^{\mu }\right\rangle _{\mathrm{J}}^{(0)}+\frac{\left( -2r\right)
^{\delta _{3\mu }-2}w^{6}}{\pi ^{2}a^{5}w_{0}^{2}}\sum_{(p)}p^{2-\delta
_{3\mu }}\frac{W_{\nu _{1}}^{(\mu )}(pw)}{U_{\nu _{2}}^{\mathrm{(J)}}(pw_{0})%
}  \notag \\
&&\times \left[ \sum_{k=1}^{[q/2]}(-1)^{k}c_{k}\cos (2\pi k\alpha _{0})\frac{%
K_{2-\delta _{3\mu }}(2rps_{k})}{s_{k}^{2-\delta _{3\mu }}}\right.  \notag \\
&&\left. +\frac{q}{\pi }\int_{0}^{\infty }du\frac{h(q,\alpha _{0},2u)\sinh u%
}{\cosh (2qu)-\cos (q\pi )}\,\frac{K_{2-\delta _{3\mu }}(2rp\cosh u)}{\cosh
^{2-\delta _{3\mu }}u}\right] .  \label{TmuJ3}
\end{eqnarray}%
In order to discuss the effects induced by the brane, it is useful to have
the VEV in the brane-free geometry. We will denote it by $\langle T_{\mu
}^{\mu }\rangle _{\mathrm{cs}}^{\mathrm{AdS}}$. The corresponding expression
is obtained from (\ref{TmuJ3}) with $W_{\nu }(pw)=J_{\nu }(pw)$, $U_{\nu
_{2}}^{\mathrm{(J)}}(pw_{0})\rightarrow 1$ and $\sum_{(p)}=w_{0}^{2}%
\int_{0}^{\infty }dp\,p$. It is decomposed as%
\begin{equation}
\langle T_{\mu }^{\mu }\rangle _{\mathrm{cs}}^{\mathrm{AdS}}=\langle T_{\mu
}^{\mu }\rangle ^{\mathrm{AdS}}+\langle T_{\mu }^{\mu }\rangle _{\mathrm{cs}%
},  \label{Tbfree}
\end{equation}%
where $\langle T_{\mu }^{\mu }\rangle ^{\mathrm{AdS}}$ is the VEV\ in pure
AdS spacetime and the string-induced contribution is given by \cite{Bell21b}
(no summation over $\mu =0,3$)%
\begin{eqnarray}
\left\langle T_{\mu }^{\mu }\right\rangle _{\mathrm{cs}} &=&\frac{a^{-5}w^{6}%
}{\pi ^{2}\left( -2r\right) ^{2-\delta _{3\mu }}}\int_{0}^{\infty
}dp\,p^{3-\delta _{3\mu }}W_{0,\nu _{1}}^{(\mu )}(pw)  \notag \\
&&\times \left[ \sum_{k=1}^{[q/2]}(-1)^{k}c_{k}\cos (2\pi k\alpha _{0})\frac{%
K_{2-\delta _{3\mu }}(2rps_{k})}{s_{k}^{2-\delta _{3\mu }}}\right.  \notag \\
&&\left. +\frac{q}{\pi }\int_{0}^{\infty }du\frac{h(q,\alpha _{0},2u)\sinh u%
}{\cosh (2qu)-\cos (q\pi )}\,\frac{K_{2-\delta _{3\mu }}(2rp\cosh u)}{\cosh
^{2-\delta _{3\mu }}u}\right] ,  \label{Tcs}
\end{eqnarray}%
where%
\begin{eqnarray}
W_{0,\nu _{1}}^{(0)}(pw) &=&J_{\nu _{1}}^{2}(pw)+J_{\nu _{2}}^{2}(pw),
\notag \\
W_{0,\nu _{1}}^{(3)}(pw) &=&-p^{2}W_{0,\nu _{1}}^{(0)}(pw)+2ma\frac{p}{w}%
J_{\nu _{1}}(pw)J_{\nu _{2}}(pw).  \label{W0}
\end{eqnarray}%
Note that this contribution is the same for $s=+1$ and $s=-1$.

The VEV of the energy-momentum tensor is decomposed as%
\begin{eqnarray}
\left\langle T_{\mu }^{\mu }\right\rangle _{\mathrm{J}} &=&\langle T_{\mu
}^{\mu }\rangle ^{\mathrm{AdS}}+\langle T_{\mu }^{\mu }\rangle _{\mathrm{cs}%
}+\langle T_{\mu }^{\mu }\rangle _{\mathrm{b,J}}^{(0)}+\frac{a^{-5}w^{6}}{%
\pi ^{2}\left( -2r\right) ^{2-\delta _{3\mu }}}  \notag \\
&&\times \left[ \sum_{k=1}^{[q/2]}(-1)^{k}c_{k}\cos (2\pi k\alpha _{0})\frac{%
f_{\mathrm{(J)}}^{(\mu )}(w_{0},w,2rps_{k})}{s_{k}^{2-\delta _{3\mu }}}%
\right.  \notag \\
&&\left. +\frac{q}{\pi }\int_{0}^{\infty }du\frac{h(q,\alpha _{0},2u)\sinh u%
}{\cosh (2qu)-\cos (q\pi )}\,\frac{f_{\mathrm{(J)}}^{(\mu
)}(w_{0},w,2rp\cosh u)}{\cosh ^{2-\delta _{3\mu }}u}\right] .  \label{TmuJ4}
\end{eqnarray}%
where the functions $f_{\mathrm{(J)}}^{(\mu )}(w_{0},w,\gamma )$ are defined
in Appendix \ref{sec:App1} and $\langle T_{\mu }^{\mu }\rangle _{\mathrm{b,J}%
}^{(0)}=\langle T_{\mu }^{\mu }\rangle _{\mathrm{J}}^{(0)}-\langle T_{\mu
}^{\mu }\rangle ^{\mathrm{AdS}}$ is the contribution of the brane in the
geometry where the cosmic string is absent. The expression for that
contribution in the case $s=1$ has been obtained in \cite{Eliz13} for
general spatial dimension (the VEV of the energy-momentum tensor for scalar
and vector fields in both single and two-brane geometries were investigated
in \cite{Knap04}). Specifying for the case $D=4$ and generalizing for $s=-1$%
, the corresponding formulas read (no summation over $\mu $)%
\begin{equation}
\langle T_{\mu }^{\mu }\rangle _{\mathrm{b,J}}^{(0)}=-\frac{4^{\delta _{3\mu
}-2}w^{6}}{\pi ^{2}a^{5}}\int_{0}^{\infty }dp\,p^{5}F_{\mathrm{(J)}}^{(\mu
)}(pw_{0},pw),  \label{TbJ}
\end{equation}%
with the functions%
\begin{eqnarray}
F_{\mathrm{(R)}}^{(\mu )}(x,y) &=&\frac{I_{\nu _{2}}(x)}{K_{\nu _{2}}(x)}%
\left[ K_{\nu _{2}}^{2}(y)-K_{\nu _{1}}^{2}(y)\right] ,\;\mu =0,1,2,4,
\notag \\
F_{\mathrm{(R)}}^{(3)}(x,y) &=&\frac{I_{\nu _{2}}(x)}{K_{\nu _{2}}(x)}\left[
K_{\nu _{1}}^{2}(y)-K_{\nu _{2}}^{2}(y)+\frac{2sma}{y}K_{\nu _{1}}(y)K_{\nu
_{2}}(y)\right] .  \label{F3R}
\end{eqnarray}%
in the R-region and%
\begin{eqnarray}
F_{\mathrm{(L)}}^{(\mu )}(x,y) &=&\frac{K_{\nu _{1}}(x)}{I_{\nu _{1}}(x)}%
\left[ I_{\nu _{1}}^{2}(y)-I_{\nu _{2}}^{2}(y)\right] ,\;\mu =0,1,2,4,
\notag \\
F_{\mathrm{(L)}}^{(3)}(x,y) &=&\frac{K_{\nu _{1}}(x)}{I_{\nu _{1}}(x)}\left[
I_{\nu _{2}}^{2}(y)-I_{\nu _{1}}^{2}(y)+\frac{2sma}{y}I_{\nu _{1}}(y)I_{\nu
_{2}}(y)\right] ,  \label{F3L}
\end{eqnarray}%
in the L-region.

By making use of the representations for the functions $f_{\mathrm{(J)}%
}^{(\mu )}(w_{0},w,\gamma )$, given in Appendix \ref{sec:App1}, the final
expression for the VEV of the energy-momentum tensor is decomposed as (no
summation over $\mu $)%
\begin{equation}
\left\langle T_{\mu }^{\mu }\right\rangle _{\mathrm{J}}=\langle T_{\mu
}^{\mu }\rangle ^{\mathrm{AdS}}+\langle T_{\mu }^{\mu }\rangle _{\mathrm{cs}%
}+\langle T_{\mu }^{\mu }\rangle _{\mathrm{b,J}},  \label{Tmudec}
\end{equation}%
where the contribution induced by the brane is given by the formula%
\begin{equation}
\langle T_{\mu }^{\mu }\rangle _{\mathrm{b,J}}=-\frac{w^{6}}{\pi ^{2}a^{5}}%
\int_{0}^{\infty }dp\,p^{5}F_{\mathrm{(J)}}^{(\mu )}(pw_{0},pw)H^{(\mu
)}(q,\alpha _{0},2pr).  \label{TmuJ5}
\end{equation}%
Here, the functions $F_{\mathrm{(J)}}^{(\mu )}(pw_{0},pw)$ in the R- and
L-regions ($\mathrm{J}=\mathrm{R}$ and $\mathrm{J}=\mathrm{L}$,
respectively) are given by (\ref{F3R}), (\ref{F3L}) and we have defined the
functions%
\begin{eqnarray}
H^{(\mu )}(q,\alpha _{0},x) &=&H_{2}(q,\alpha _{0},x),\;\mu =0,1,4,  \notag
\\
H^{(2)}(q,\alpha _{0},x) &=&H_{1}(q,\alpha _{0},x)-3H_{2}(q,\alpha _{0},x),
\notag \\
H^{(3)}(q,\alpha _{0},x) &=&H_{1}(q,\alpha _{0},x),  \label{Hmu}
\end{eqnarray}%
with $H_{n}(q,\alpha _{0},x)$ from (\ref{Hnq}). The contribution of the $k=0$
term in (\ref{Hnq}) to (\ref{TmuJ5}) presents the VEV induced by the brane
in the geometry where the cosmic string is absent. It is given by the
expression (\ref{TbJ}). For points outside the defect core and the brane,
the last two contributions are finite and the renormalization is needed for
the part $\langle T_{\mu }^{\mu }\rangle ^{\mathrm{AdS}}$ only. From the
maximal symmetry of the background geometry for that part it follows that
for the renormalized VEV we have $\langle T_{\mu }^{\nu }\rangle ^{\mathrm{%
AdS}}=\mathrm{const}\cdot \delta _{\mu }^{\nu }$. Comparing the expressions
for the energy-momentum tensor in the R- and L-regions, we can see that the
brane-induced contribution in the L-region is obtained from the
corresponding one in the R-region by the replacements $K\rightarrow I$, $%
I\rightarrow K$, of the modified Bessel functions, and $\nu
_{1,2}\rightarrow \nu _{2,1}$ in their orders. The special cases of the
general formula (\ref{TmuJ5}) in the absence of the magnetic flux or planar
angle deficit are obtained by using the corresponding expressions (\ref%
{Hnqalf0}) and (\ref{Hnq1}) in (\ref{Hmu}).

The parts $\langle T_{\mu }^{\nu }\rangle _{\mathrm{cs}}$ and $\langle
T_{\mu }^{\nu }\rangle _{\mathrm{b,J}}$ separately obey the trace relations $%
\langle T_{\mu }^{\mu }\rangle _{\mathrm{cs}}=$ $sm\langle \bar{\psi}\psi
\rangle _{\mathrm{cs}}$ and $\langle T_{\mu }^{\mu }\rangle _{\mathrm{b,J}}=$
$sm\langle \bar{\psi}\psi \rangle _{\mathrm{b,J}}$. In particular, these
tensors are traceless for a massless fermionic field. Moreover, it can be
verified that the covariant conservation equation, $\nabla _{\mu }\langle
T^{\mu \nu }\rangle =0$, is also satisfied by the separate terms in (\ref%
{Tmudec}). In the problem at hand, it is reduced to the following two
relations
\begin{eqnarray}
&&\partial _{r}(r\langle T_{1}^{1}\rangle )-\langle T_{2}^{2}\rangle =0\ ,
\notag \\
&& w\partial _{w}\langle T_{3}^{3}\rangle -5\langle T_{3}^{3}\rangle
+\langle T_{\mu }^{\mu }\rangle =0\ ,  \label{conseq}
\end{eqnarray}%
between the separate components. By taking into account that $\langle
T_{1}^{1}\rangle =\langle T_{0}^{0}\rangle $, we see that the first relation
in (\ref{conseq}) coincides with (\ref{T2}).

\begin{figure}[tbph]
\begin{center}
\begin{tabular}{cc}
\epsfig{figure=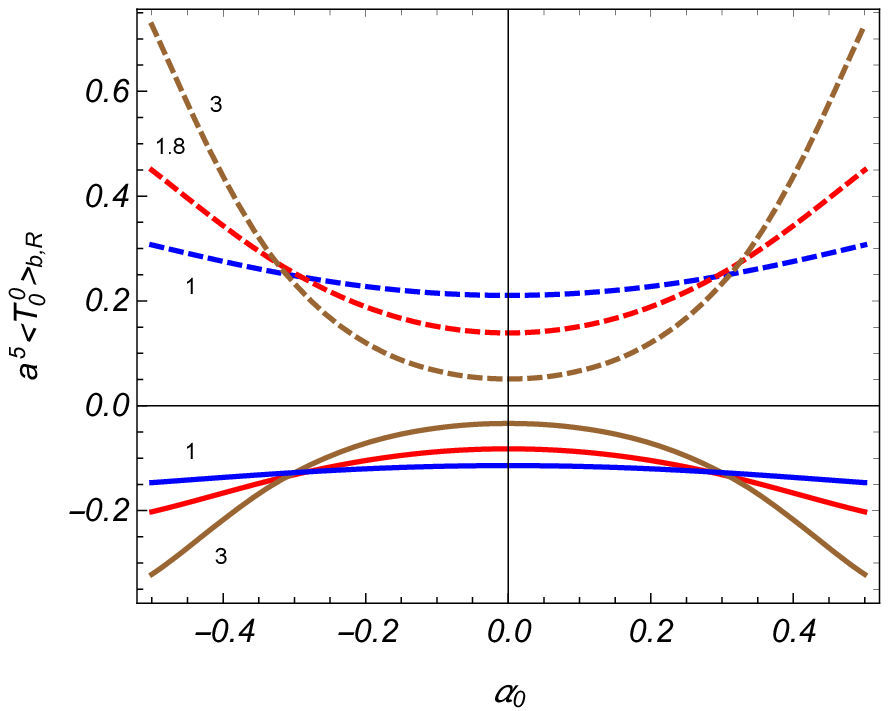,width=7.cm,height=5.5cm} & \quad %
\epsfig{figure=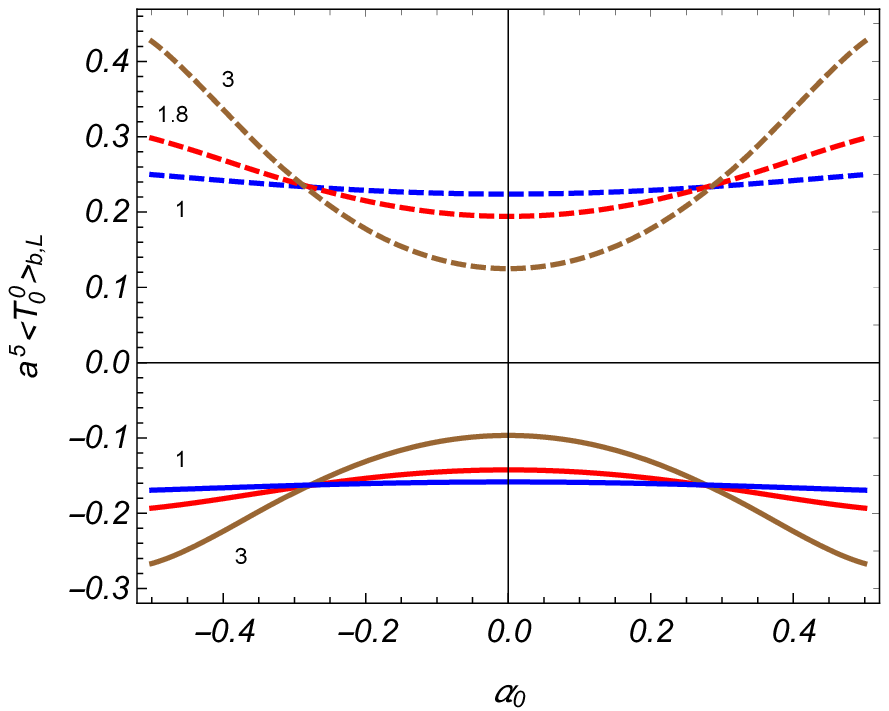,width=7.cm,height=5.5cm}%
\end{tabular}%
\end{center}
\caption{The brane-induced energy density as a function of the parameter $%
\protect\alpha _{0}$. The left and right panels correspond to the R- (with $%
w/w_{0}=1.5$) and L- (with $w/w_{0}=0.75$) regions and the numbers near the
curves are the values for $q$. The full and dashed curves are plotted for $%
s=+1$ and $s=-1$, respectively, and for $ma=1$, $r/w_{0}=0.25$.}
\label{figT00alf}
\end{figure}

The VEV of the energy-momentum tensor is an even periodic function of the
magnetic flux with the period equal to the flux quantum. That is expressed
in terms of the dependence on the parameter $\alpha _{0}$. That dependence
for the brane-induced vacuum energy density in the R-region (for $%
w/w_{0}=1.5 $, left panel) and L-region (for $w/w_{0}=0.75$, right panel) is
displayed in Figure \ref{figT00alf} for different values of the parameter $q$
(the numbers near the curves). The full and dashed curves correspond to the
fields with $s=+1$ and $s=-1$. The graphs are plotted for $ma=1$ and $%
r/w_{0}=0.25$. Similar to the case of the FC, we see that the dependence on
the magnetic flux is stronger for higher values of the planar angle angle
deficit.

\subsection{Asymptotic analysis and numerical results}

In order to clarify the behavior of the vacuum energy-momentum tensor we
consider special cases and asymptotics.

\subsubsection{Minkowskian limit}

For large values of the curvature radius $a$ the orders of the modified
Bessel functions in the integrand of (\ref{TbJ}) are large and we use the
corresponding uniform asymptotic expansions. It can be seen that in the
leading order we get
\begin{equation}
F_{\mathrm{(J)}}^{(\mu )}(x,p)\approx -\frac{2s}{pw}F_{\mathrm{(J)}%
}(pw_{0},pw),\;\mu =0,1,2,4,  \label{FJmuM}
\end{equation}%
where the asymptotic expression for $F_{\mathrm{(J)}}(pw_{0},pw)$ is given
by (\ref{FJM}). Substituting the asymptotic expression (\ref{FJmuM}) in (\ref%
{TmuJ5}), we see that the leading term in the expansion of $\langle T_{\mu
}^{\mu }\rangle _{\mathrm{b,J}}$ does not depend on $a$ and it coincides
with the corresponding VEV induced by a planar boundary in the
(4+1)-dimensional Minkowski spacetime with a cosmic string. Denoting the
latter by $\langle T_{\mu }^{\nu }\rangle _{\mathrm{b}}^{\mathrm{(M)}}$, for
the components with $\mu \neq 3$ one obtains (no summation over $\mu $)%
\begin{equation}
\langle T_{\mu }^{\mu }\rangle _{\mathrm{b}}^{\mathrm{(M)}}=\frac{m}{\pi ^{2}%
}\int_{m}^{\infty }dx\,(x^{2}-m^{2})e^{-2x|y-y_{0}|}\left( m-sx\right)
H^{(\mu )}(q,\alpha _{0},2r\sqrt{x^{2}-m^{2}}).  \label{TbJM}
\end{equation}%
Of course, in the Minkowski bulk the VEVs are symmetric with respect to the
boundary. Combining (\ref{TbJM}) result with the trace relation $\langle
T_{\mu }^{\mu }\rangle _{\mathrm{b}}^{\mathrm{(M)}}=sm\langle \bar{\psi}\psi
\rangle _{\mathrm{b}}^{\mathrm{(M)}}$ and by taking into account (\ref{FCbMi}%
) we can see that $\langle T_{3}^{3}\rangle _{\mathrm{b}}^{\mathrm{(M)}}=0$.
This result could also be directly obtained from the continuity equation $%
\nabla _{\mu }\langle T^{\mu \nu }\rangle _{\mathrm{b}}^{\mathrm{(M)}}=0$.

\subsubsection{Massless field}

For a massless fermionic field one has $\nu _{1}=-\nu _{2}=-s/2$, and by
taking into account the property $K_{\nu }(z)=K_{-\nu }(z)$ for the
Macdonald function, we see that the brane-induced VEV of the energy-momentum
tensor vanishes in the R-region. This result could be directly obtained
based on the conformal relation with the problem of a single boundary on the
Minkowski bulk with a cosmic string.\footnote{%
In Ref. \cite{Beze12b} the authors have shown that the boundary-induced
contribution in the VEV of the energy-momentum tensor for a massless
fermionic field is zero for a planar boundary in (3+1)-dimensional Minkowski
spacetime with a straight cosmic string perpendicular to the boundary.} For
the L-region one has $F_{\mathrm{(L)}}^{(\mu )}(x,y)=-F_{\mathrm{(L)}%
}^{(3)}(x,y)=2/[y(se^{2x}+1)]$, $\mu \neq 3$, and the VEV is presented as
(no summation over $\mu $)%
\begin{equation}
\langle T_{\mu }^{\mu }\rangle _{\mathrm{b,L}}=-\frac{2(-1)^{\delta _{3\mu }}%
}{\pi ^{2}a^{5}}\int_{0}^{\infty }dx\,\frac{x^{4}H^{(\mu )}(q,\alpha
_{0},2xr/w)}{se^{2xw_{0}/w}+1}.  \label{TbJm0}
\end{equation}%
This result is conformally related to the corresponding formula for a cosmic
string in the Minkowski bulk with the line element (\ref{ds2M}) and with two
planar boundaries located at $w=0$ and $w=w_{0}$. In the case of a massless
field for the brane-free part one has \cite{Bell21b} (no summation over $\mu
$)%
\begin{equation}
\left\langle T_{\mu }^{\mu }\right\rangle _{\mathrm{cs}}=\frac{%
3h_{5}(q,\alpha _{0})}{32\pi ^{2}r_{p}^{5}}\,,  \label{Tllm0}
\end{equation}%
for $\mu =0,1,3,4$ and $\langle T_{2}^{2}\rangle _{\mathrm{cs}}=-4\langle
T_{0}^{0}\rangle _{\mathrm{cs}}$.

\subsubsection{Asymptotics with respect to the brane location}

Here we consider the asymptotic behavior of the vacuum energy-momentum
tensor with respect to the ratio $w/w_{0}$. In the R-region and for large
values of this ratio, the dominant contribution to the integral (\ref{TmuJ5}%
) comes from the region where $pw\lesssim 1$. In that region $pw_{0}$ is
small and we expand the corresponding modified Bessel functions. To the
leading order this gives (no summation over $\mu $)%
\begin{eqnarray}
\langle T_{\mu }^{\mu }\rangle _{\mathrm{b,R}} &\approx &-\frac{2\left(
w_{0}/2w\right) ^{\nu _{2}+|\nu _{2}|}}{\pi ^{2}a^{5}\Gamma \left( \nu
_{2}+1\right) \Gamma (|\nu _{2}|)}\int_{0}^{\infty }dx\,x^{5+\nu _{2}+|\nu
_{2}|}H^{(\mu )}(q,\alpha _{0},2xr/w)  \notag \\
&&\times \left\{ (-1)^{\delta _{\mu 3}}\left[ K_{\nu _{2}}^{2}(x)-K_{\nu
_{1}}^{2}(x)\right] +\delta _{\mu 3}\frac{2sma}{x}K_{\nu _{1}}(x)K_{\nu
_{2}}(x)\right\} .  \label{TbRlargew}
\end{eqnarray}%
For fixed values of $w$ and $r$ and for small $w_{0}$ this asymptotic
describes the situation where the brane is close to the AdS boundary.
Similar to the case of the FC, for fields with $s=1$ and $s=-1$, $ma>1/2$
the brane-induced VEV vanishes as $w_{0}^{2ma+s}$. In the case $s=-1$, $%
ma<1/2$ the VEV $\langle T_{\mu }^{\mu }\rangle _{\mathrm{b,R}}$ tends to a
nonzero limiting value in the limit $w_{0}\rightarrow 0$. Fixing $w_{0}$ and
$r$ and for large values of $w$, we have the situation where the observation
point is close to the horizon. In this case we can additionally expand the
function $H^{(\mu )}(q,\alpha _{0},2xr/w)$ for small values of the ratio $%
r/w $. The latter corresponds to small proper distances from the string
compared with the AdS curvature radius. The corresponding asymptotic is
discussed below.

In the L-region and for small values of the ratio $w/w_{0}$ we introduce a
new integration variable $x=pw$ and expand the integrand with respect to $%
w/w_{0}$. The leading term is expressed as%
\begin{eqnarray}
\langle T_{\mu }^{\mu }\rangle _{\mathrm{b,L}} &\approx &-\frac{%
2^{1-2ma}s\left( w/w_{0}\right) ^{5+2ma}}{\pi ^{2}a^{5}\Gamma ^{2}\left(
ma+1/2\right) }\left( \frac{-1}{2ma+1}\right) ^{\delta _{\mu 3}}  \notag \\
&&\times \int_{0}^{\infty }dx\,x^{4+2ma}\frac{K_{\nu _{1}}(x)}{I_{\nu
_{1}}(x)}H^{(\mu )}(q,\alpha _{0},2xr/w_{0}).  \label{TbLnearB}
\end{eqnarray}%
Again, we can consider two situations. The first one corresponds to large
values of $w_{0}$ for fixed $w$ and $r$ (the brane is close to horizon) and
we can further expand the function $H^{(\mu )}(q,\alpha _{0},2xr/w_{0})$ for
small $r/w_{0}$. The second situation corresponds to small values of $w$ for
fixed $w_{0}$ and $r$ (the observation point is close to the AdS boundary).
As seen, for fixed $r/w_{0}>0$, the brane-induced VEV $\langle T_{\mu }^{\mu
}\rangle _{\mathrm{b,L}}$ vanishes on the AdS boundary like $w^{5+2ma}$. A
similar behavior is exhibited by the brane-free part as well (see \cite%
{Bell21b}).

Now let us consider the brane-induced VEV (\ref{TmuJ5}) near the brane. All
the terms in (\ref{TmuJ5}), except the $k=0$ term in the summation over $k$,
are finite on the brane. The $k=0$ term corresponds to the brane-induced VEV
in the problem where the cosmic string is absent and for a massive field it
diverges on the brane. This means that near the brane the VEV (\ref{TmuJ5})
is dominated by that term. For points near the brane the main contribution
to the corresponding integral comes from the region where $p$ is large. By
using the asymptotic expressions for the modified Bessel function with large
arguments \cite{Abra}, we can see that $F_{\mathrm{(R)}}^{(\mu
)}(pw_{0},pw)\approx smae^{-2p|w-w_{0}|}/(pw)^{2}$ for $\mu \neq 3$. Hence,
to the leading order, for the components with $\mu \neq 3$ one gets (no
summation over $\mu $)%
\begin{equation}
\langle T_{\mu }^{\mu }\rangle _{\mathrm{J}}\approx -\frac{3sm}{32\pi
^{2}a^{4}\left( w_{0}/w-1\right) ^{4}}.  \label{TmuJnear}
\end{equation}%
The asymptotic expression for the stress $\langle T_{3}^{3}\rangle _{\mathrm{%
J}}$ is most conveniently obtained by using the second equation in (\ref%
{conseq}):
\begin{equation}
\langle T_{3}^{3}\rangle _{\mathrm{J}}\approx \frac{sm\,}{8\pi
^{2}a^{4}\left( w_{0}/w-1\right) ^{3}}.  \label{T33Jnear}
\end{equation}%
As seen, the leading term is symmetric with respect to the brane for the
components with $\mu \neq 3$ and has opposite signs for the normal stress $%
\langle T_{3}^{3}\rangle _{\mathrm{J}}$. As it has been mentioned above,
near the brane one has $1-w_{0}/w\approx (y-y_{0})/a$ and the leading term
in (\ref{TmuJnear}) coincides with the corresponding result for a planar
boundary in the Minkowski bulk with a cosmic string where the distance from
the boundary is given by $|y-y_{0}|$. Note that in the Minkowski bulk the
normal stress $\langle T_{3}^{3}\rangle _{\mathrm{J}}$ is zero and the
nonzero effect in the right-hand side of (\ref{T33Jnear}) is an effect
induced by the background curvature. For a massless field the brane-induced
VEV $\langle T_{\mu }^{\mu }\rangle _{\mathrm{b,J}}$ is finite on the brane
and the same is the case for the total VEV\ $\left\langle T_{\mu }^{\mu
}\right\rangle _{\mathrm{J}}$. In this case the brane-induced part is zero
in the R-region and the limiting value $\langle T_{\mu }^{\mu }\rangle _{%
\mathrm{b,L}}|_{w=w_{0}}$ for the L-region is directly obtained from (\ref%
{TbJm0}) putting in the integrand $w=w_{0}$.

\begin{figure}[tbph]
\begin{center}
\begin{tabular}{cc}
\epsfig{figure=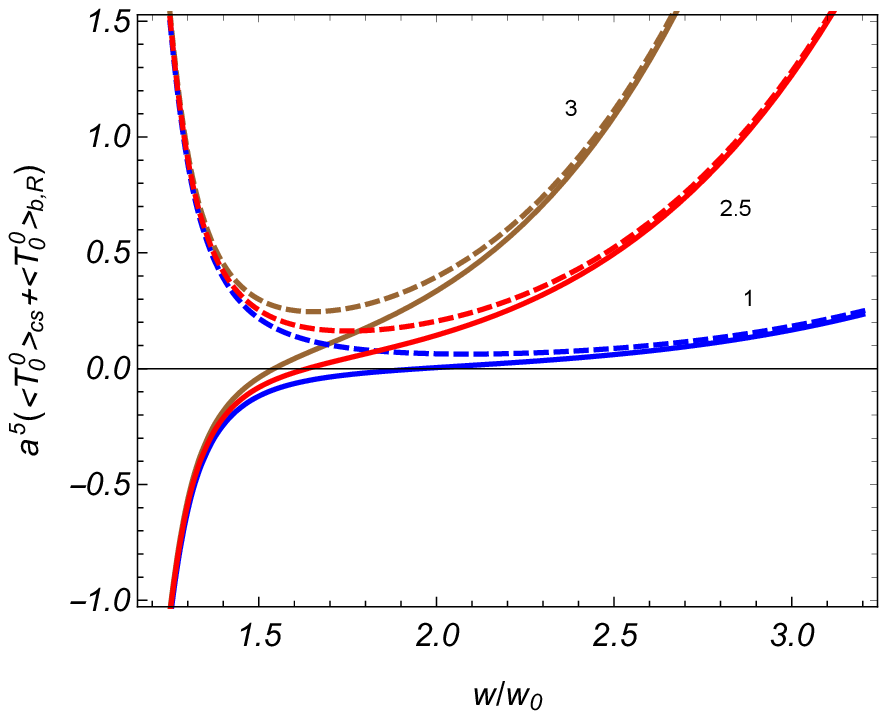,width=7.cm,height=5.5cm} & \quad %
\epsfig{figure=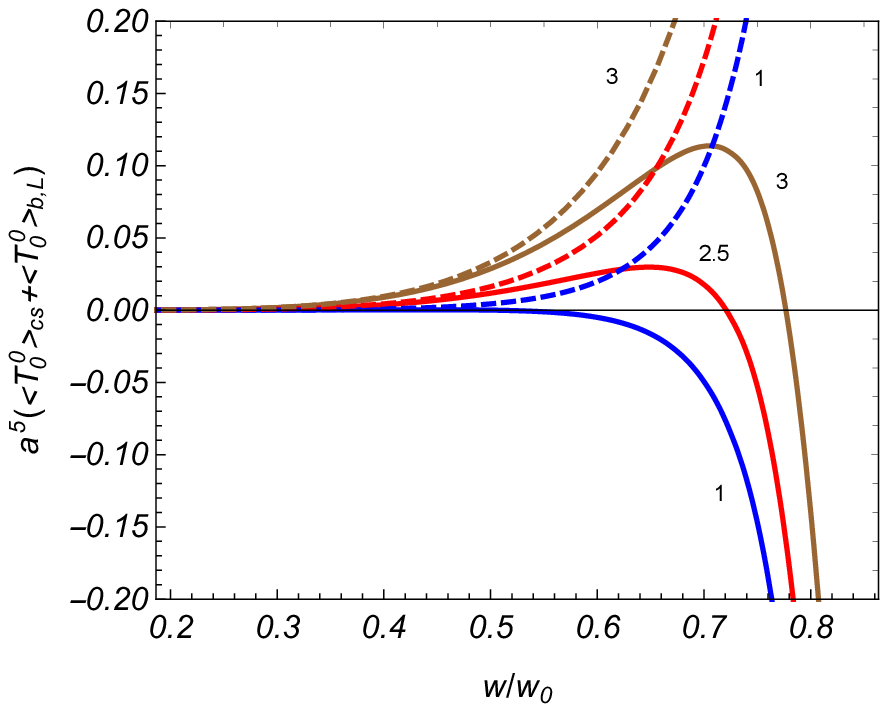,width=7.cm,height=5.5cm}%
\end{tabular}%
\end{center}
\caption{The dependence of the cosmic string and brane-induced contribution
in the VEV of the energy density on the ratio $w/w_{0}$ for different values
of the parameter $q$ (numbers near the curves). The full and dashed curves
correspond to the fields with $s=+1$ and $s=-1$. The graphs are plotted for $%
ma=1$, $\protect\alpha _{0}=0.3$, $r/w_{0}=0.75$.}
\label{figT00Rw}
\end{figure}

In Figure \ref{figT00Rw} we present the combined effects of the cosmic
string and brane in the VEV of the vacuum energy density, considered as a
function of the ratio $w/w_{0}$. The left and right panels are plotted for
the R- and L-regions and the numbers near the curves are the values of the
parameter $q$. The full and dashed curves correspond to the fields with $%
s=+1 $ and $s=-1$, respectively, and the numbers near the curves are the
values of $q$. For the values of the parameters we have taken $ma=1$, $%
\alpha _{0}=0.3$, $r/w_{0}=0.75$. The vacuum energy density is dominated by
the brane-induced contribution near the brane and by the brane-free part
near the horizon. It can be either positive or negative, depending on the
distance from the brane.

\subsubsection{Small and large distances from the cosmic string}

It remains to consider the asymptotics near the cosmic string and at large
distances from it. For points away from the brane and for $2|\alpha
_{0}|<1-1/q$, the brane-induced contribution is finite on the string ($r=0$)
and we can directly put $r=0$ in the integrand of (\ref{TmuJ5}). This leads
to the result (no summation over $\mu $)%
\begin{equation}
\langle T_{\mu }^{\mu }\rangle _{\mathrm{b,J}}|_{r=0}=\left[
1+2h_{0}(q,\alpha _{0})\right] \langle T_{\mu }^{\mu }\rangle _{\mathrm{b,J}%
}^{(0)},  \label{TmuJr0}
\end{equation}%
where the brane-induced contribution $\langle T_{\mu }^{\mu }\rangle _{%
\mathrm{b,J}}^{(0)}$ in the geometry without cosmic string is given by (\ref%
{TbJ}). As it has been already explained before, $1+2h_{0}(q,\alpha _{0})=0$
in the range $2|\alpha _{0}|<1-1/q$ and in that range of the parameters the
brane-induced term $\langle T_{\mu }^{\mu }\rangle _{\mathrm{b,J}}$ vanishes
on the string. In the special case $2|\alpha _{0}|=1-1/q$ in (\ref{TmuJr0})
we have $1+2h_{0}(q,\alpha _{0})=q/2$. For $2|\alpha _{0}|>1-1/q$ and near
the cosmic string the main contribution in (\ref{TmuJ5}) comes from the term
containing the integral over $u$. The integral is dominated by the region
with large values of $u$. By using the corresponding asymptotic expressions,
in the way similar to that for the FC, we get (no summation over $\mu $)%
\begin{equation}
\langle T_{\mu }^{\mu }\rangle _{\mathrm{b,J}}\approx -\frac{qa^{-5}B_{\mu
}\left( r/2w\right) ^{2\alpha _{q}-1}}{8\pi ^{2}\Gamma \left( 5/2+\alpha
_{q}\right) \Gamma (1/2+\alpha _{q})}\int_{0}^{\infty }dp\,p^{4+2\alpha
_{q}}F_{\mathrm{(J)}}^{(\mu )}(pw_{0}/w,p),  \label{TmuJnearSt}
\end{equation}%
with the notations%
\begin{equation}
B_{\mu }=\frac{1}{2},\;\mu =0,1,4,\;B_{2}=\alpha _{q},\;B_{3}=\frac{3}{2}%
+\alpha _{q}.  \label{Amu}
\end{equation}%
In the limit $\alpha _{q}\rightarrow 1/2$ the result (\ref{TmuJnearSt}) is
reduced to (\ref{TmuJr0}). As regards the brane-free contribution $\langle
T_{\mu }^{\mu }\rangle _{\mathrm{cs}}$, near the string the effect of the
mass is weak and to the leading order it coincides with that for a massless
field. In that region the influence of the background gravitational field on
the string induced effects is weak and the VEV $\left\langle T_{\mu }^{\mu
}\right\rangle _{\mathrm{cs}}$ behaves as $(w/r)^{5}$ (see (\ref{Tllm0})).

In the discussion of the asymptotic for the cosmic string-induced
contribution at large distances from the core it is more convenient to use
the representation (\ref{TmuJ3}). In the R-region one has $%
\sum_{(p)}=w_{0}^{2}\int_{0}^{\infty }dp\,p$ and the dominant contribution
to the integral in (\ref{TmuJ3}) comes from the integration range $p\lesssim
1/r$. In that range the arguments of the Bessel and Neumann functions, $%
w_{0}p$ and $wp$, are small in the region under consideration. By employing
the asymptotics for those functions \cite{Abra}, the integral over $p$ is
evaluated by making use of the formula (\ref{IntK}). For $s=1$ and $s=-1$, $%
ma<1/2$, to the leading order, for the components with $\mu =0,1,4$ one
finds (no summation over $\mu $)%
\begin{equation}
\left\langle T_{\mu }^{\mu }\right\rangle _{\mathrm{R}}\approx \left\langle
T_{\mu }^{\mu }\right\rangle _{\mathrm{R}}^{(0)}+\frac{\left( 3+2sma\right)
\left( 1+2sma\right) }{\pi ^{2}a^{5}\left( 2r/w\right) ^{5+2sma}}%
h_{5+2sma}(q,\alpha _{0}).  \label{Tmularger}
\end{equation}%
For the contributions in the remaining components, induced by the cosmic
string, we have%
\begin{eqnarray}
\left\langle T_{2}^{2}\right\rangle _{\mathrm{R}}-\left\langle
T_{2}^{2}\right\rangle _{\mathrm{R}}^{(0)} &\approx &-2(2+sma)\left[
\left\langle T_{0}^{0}\right\rangle _{\mathrm{R}}-\left\langle
T_{0}^{0}\right\rangle _{\mathrm{R}}^{(0)}\right] ,  \notag \\
\left\langle T_{3}^{3}\right\rangle _{\mathrm{R}}-\left\langle
T_{3}^{3}\right\rangle _{\mathrm{R}}^{(0)} &\approx &\left[
1+2sma(w_{0}/w)^{1+2sma}\right] \left[ \left\langle T_{0}^{0}\right\rangle _{%
\mathrm{R}}-\left\langle T_{0}^{0}\right\rangle _{\mathrm{R}}^{(0)}\right] .
\label{T33Rlarger}
\end{eqnarray}%
In the case $s=-1$, $ma>1/2$ the asymptotic at large distances has the form
(no summation over $\mu $)%
\begin{equation}
\left\langle T_{\mu }^{\mu }\right\rangle _{\mathrm{R}}\approx \left\langle
T_{\mu }^{\mu }\right\rangle _{\mathrm{R}}^{(0)}+\frac{\left(
4m^{2}a^{2}-1\right) \left( w_{0}/w\right) ^{4ma-2}}{\pi
^{2}a^{5}(2r/w)^{3+2ma}}h_{3+2ma}(q,\alpha _{0}),  \label{Tmularger2}
\end{equation}%
for $\mu =0,1,4$ and%
\begin{eqnarray}
\left\langle T_{2}^{2}\right\rangle _{\mathrm{R}}-\left\langle
T_{2}^{2}\right\rangle _{\mathrm{R}}^{(0)} &\approx &-2(1+ma)\left[
\left\langle T_{0}^{0}\right\rangle _{\mathrm{R}}-\left\langle
T_{0}^{0}\right\rangle _{\mathrm{R}}^{(0)}\right] ,  \notag \\
\left\langle T_{3}^{3}\right\rangle _{\mathrm{R}}-\left\langle
T_{3}^{3}\right\rangle _{\mathrm{R}}^{(0)} &\approx &\left[
2ma(w/w_{0})^{2ma-1}-1\right] \left[ \left\langle T_{0}^{0}\right\rangle _{%
\mathrm{R}}-\left\langle T_{0}^{0}\right\rangle _{\mathrm{R}}^{(0)}\right] ,
\label{T33Rlarger2}
\end{eqnarray}%
for the remaining stresses. As we see in the case $s=-1$ the decay of the
topological contributions is more slowly.

In the L-region the spectrum of the quantum number $p$ is discrete and the
decrease of the string-induced contributions is exponential, like $%
e^{-2rp_{1}/w_{0}}$ for $1\leq q<2$ and as $e^{-2rp_{1}s_{1}/w_{0}}$ for $%
q\geq 2$, where $p_{1}/w_{0}$ is the lowest positive eigenvalue for $p$.

\begin{figure}[tbph]
\begin{center}
\begin{tabular}{cc}
\epsfig{figure=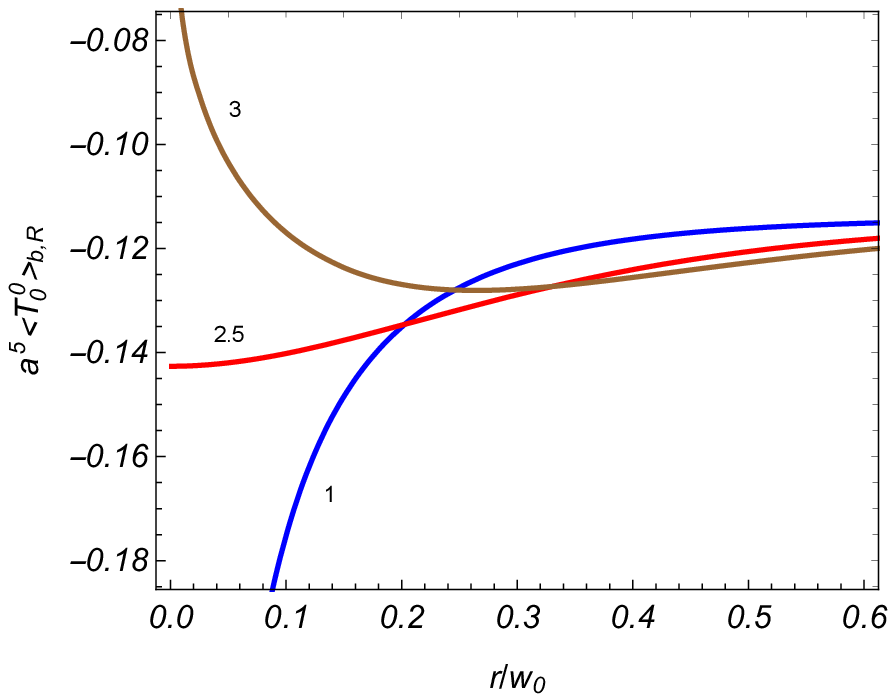,width=7.cm,height=5.5cm} & \quad %
\epsfig{figure=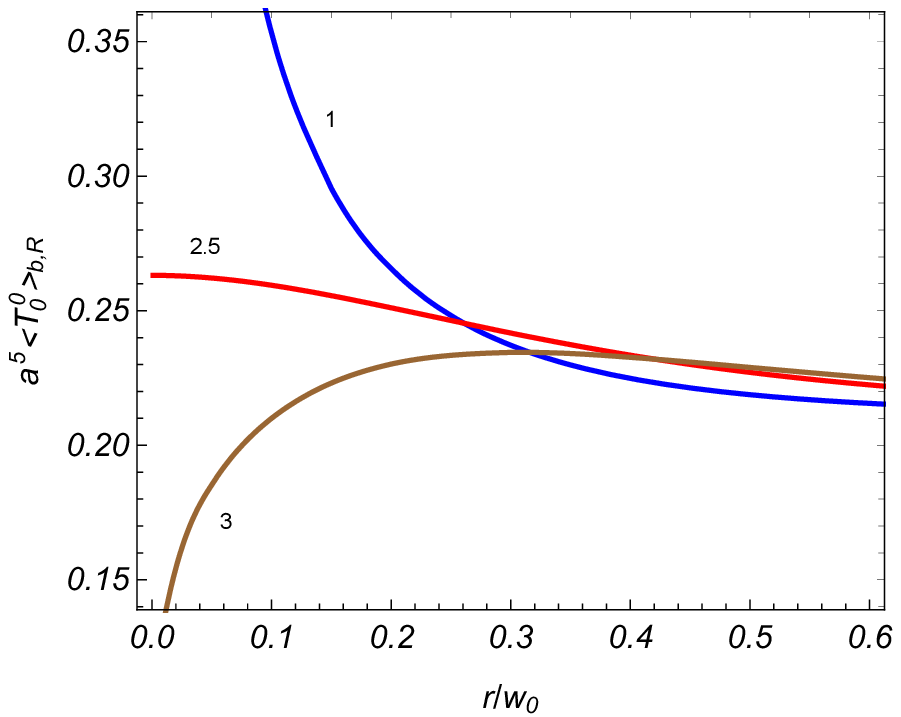,width=7.cm,height=5.5cm}%
\end{tabular}%
\end{center}
\caption{Brane-induced contribution in the VEV of the energy density versus
the distance from the cosmic string for $s=+1$ (left panel) and $s=-1$
(right panel) and for $w/w_{0}=1.5$. The numbers near the curves are the
values of the parameter $q$. The graphs are plotted for $ma=1$ and $\protect%
\alpha _{0}=0.3$.}
\label{figT00Rr}
\end{figure}

\begin{figure}[tbph]
\begin{center}
\begin{tabular}{cc}
\epsfig{figure=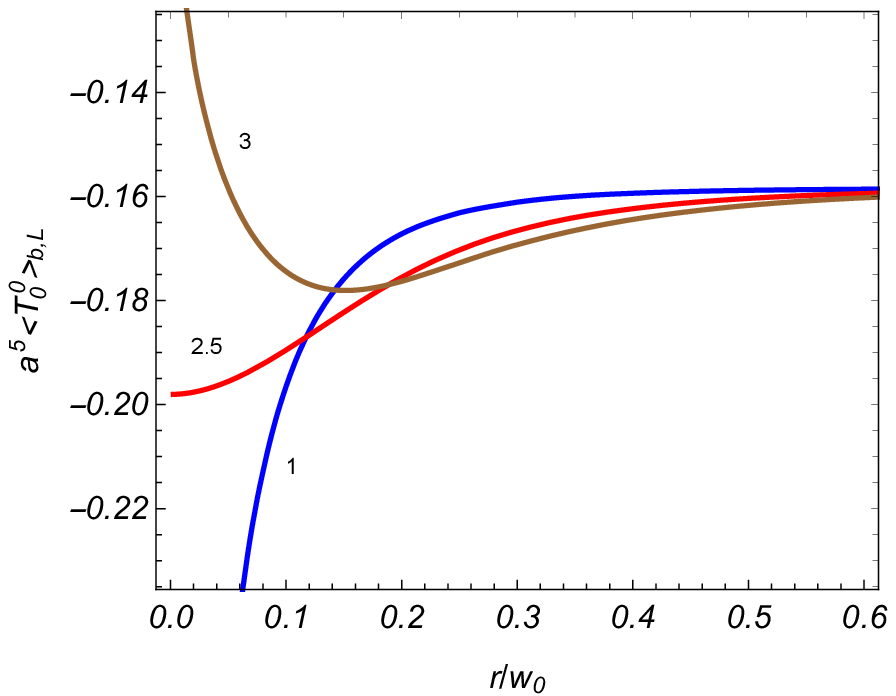,width=7.cm,height=5.5cm} & \quad %
\epsfig{figure=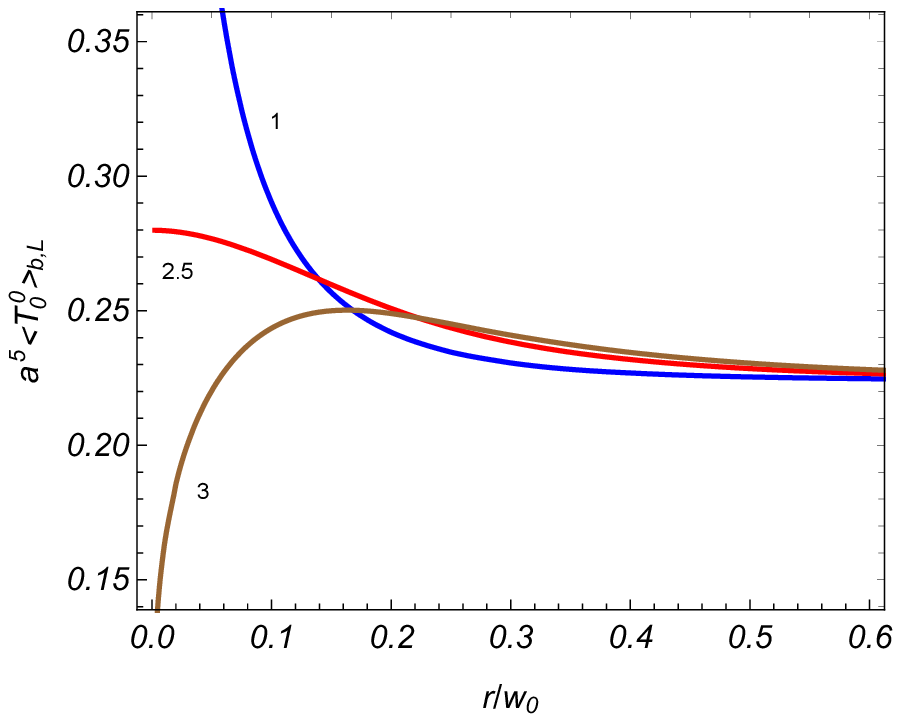,width=7.cm,height=5.5cm}%
\end{tabular}%
\end{center}
\caption{The same as in figure \protect\ref{figT00Rr} for the L-region with $%
w/w_{0}=0.75$.}
\label{figT00Lr}
\end{figure}

Figure \ref{figT00Rr} displays the dependence of the brane-induced part in
the energy density, as a function of the distance from the cosmic string, in
the R-region for $w/w_{0}=1.5$. The left and right panels correspond to the
fields with $s=+1$ and $s=-1$, respectively, and the numbers near the curves
are the values of $q$. For the remaining parameters we have taken $ma=1$ and
$\alpha _{0}=0.3$. The same graphs for $w/w_{0}=0.75$ (L-region) are
presented in Figure \ref{figT00Lr}.

\section{Vacuum densities for a field realizing the second representation
and applications in braneworlds}

\label{sec:Reps}

\subsection{Fields realizing the two representations}

We have considered a fermionic field in (4+1)-dimensional spacetime. In odd
number of spacetime dimensions the Clifford algebra for gamma matrices has
two inequivalent irreducible representations. In this section they will be
specified by $s=+1$ and $s=-1$ and the corresponding fields will be denoted
by $\psi _{(s)}$. As we will see, the parameter $s$ is identified with the
parameter $s$ in the Lagrangian density (\ref{Lsp}). In (4+1)-dimensional
flat spacetime the two sets of Dirac matrices (in the coordinates $(t,r,\phi
,w,z)$) can be taken as $\gamma _{(s)}^{(b)}=\{\gamma ^{(0)},\gamma
^{(1)},\gamma ^{(2)},\gamma ^{(3)},s\gamma ^{(4)}\}$, where the matrices $%
\gamma ^{(b)}$ are related to the matrices (\ref{gam}) as $\gamma
^{(b)}=(a/w)\delta _{\mu }^{b}$ $\gamma ^{\mu }$ and for the matrix $\gamma
^{(4)}$ one has $\gamma ^{(4)}=-r\gamma ^{(0)}\gamma ^{(1)}\gamma
^{(2)}\gamma ^{(3)}$. Introducing the corresponding curved spacetime Dirac
matrices $\gamma _{(s)}^{\mu }=(w/a)\delta _{b}^{\mu }\gamma ^{(b)}$ and the
related spin connection $\Gamma _{\mu }^{(s)}$, the Lagrangian density for
the free field $\psi _{(s)}$ is presented as $L_{(s)}=\bar{\psi}%
_{(s)}(i\gamma _{(s)}^{\mu }\mathcal{D}_{\mu }^{(s)}-m)\psi _{(s)}$ with $%
\mathcal{D}_{\mu }^{(s)}=\partial _{\mu }+\Gamma _{\mu }^{(s)}+ieA_{\mu }$.
In order to see the relation of this Lagrangian density to (\ref{Lsp}), we
introduce new fields $\psi _{(s)}^{\prime }$ in accordance with $\psi
_{(s)}^{\prime }=-\left( \gamma ^{(4)}\right) ^{1+\delta _{1s}}\psi _{(s)}$.
In terms of these fields the Lagrangian density is presented as $L_{(s)}=%
\bar{\psi}_{(s)}^{\prime }\left( i\gamma ^{\mu }\mathcal{D}_{\mu }-sm\right)
\psi _{(s)}^{\prime }$, where $\gamma ^{\mu }=\gamma _{(+1)}^{\mu }$ and $%
\mathcal{D}_{\mu }$ is the same as in (\ref{Lsp}). This shows that the
fields $\psi _{(s)}^{\prime }$ correspond to the fields with $s=+1$ and $%
s=-1 $ in the discussion of the previous sections.

Let us compare the boundary conditions for the fields $\psi _{(s)}$ and $%
\psi _{(s)}^{\prime }$. We start the discussion with the case when the
fields $\psi _{(s)}$ obey the bag boundary condition on the brane:%
\begin{equation}
(1+i\gamma _{(s)}^{\mu }n_{\mu })\psi _{(s)}=0\,,  \label{BCs}
\end{equation}%
for $w=w_{0}$. Transforming to the fields $\psi _{(s)}^{\prime }$, we can
see that they obey the condition
\begin{equation}
(1+si\gamma ^{\mu }n_{\mu })\psi _{(s)}^{\prime }=0\,,  \label{BCsp}
\end{equation}%
on $w=w_{0}$. This shows that the VEVs for the field $\psi _{(+1)}$ are
given by the formulas presented in Sections \ref{sec:FCond} and \ref%
{sec:EMT0} for the case $s=1$. The transformed field $\psi _{(-1)}^{\prime }$
obeys the equation (\ref{DiracEq}), however, the corresponding boundary
condition (\ref{BCsp}) differs from the condition (\ref{MITbc}) by the sign
of the term that contains the normal to the brane. The corresponding VEVs
are obtained in the way similar to the ones in the previous sections and are
discussed in the next subsection.

\subsection{VEVs for the second type of boundary conditions}

Let us consider the VEVs for a fermionic field $\psi $ obeying the field
equation (\ref{DiracEq}) and the boundary condition (the condition (\ref%
{BCsp}) with $s=-1$)%
\begin{equation}
(1-i\gamma ^{\mu }n_{\mu })\psi =0\,,  \label{BCm}
\end{equation}%
on the brane located at $w=w_{0}$. The complete set of mode functions still
has the form (\ref{FermMod}), where now $W_{\nu }(pw)=J_{\nu }(pw)$ in the
L-region and $W_{\nu }(pw)=G_{\nu _{1},\nu }(pw_{0},pw)$ in the R-region.
These functions obey the condition (\ref{BCm}) in the R-region and from the
boundary condition in the L-region it follows that the eigenvalues of the
quantum number $p$ in that region are roots of the equation $J_{\nu
_{2}}(pw)=0$ (compare with (\ref{peqL})). The normalization coefficients are
obtained from (\ref{Csig}) by the replacement $\nu _{2}\rightarrow \nu _{1}$
or $s\rightarrow -s$. Following the same steps as in Section \ref{sec:FCond}%
, we can see that the FC is presented in the decomposed form (\ref{FCJdec}),
where the contributions $\langle \bar{\psi}\psi \rangle ^{\mathrm{AdS}}$ and
$\langle \bar{\psi}\psi \rangle _{\mathrm{cs}}$ are given by the same
expressions and the brane-induced FC\ is given by
\begin{equation}
\langle \bar{\psi}\psi \rangle _{\mathrm{b,J}}=\frac{2w^{5}}{\pi ^{2}a^{4}}%
\int_{0}^{\infty }dp\,p^{4}F_{\mathrm{(J)}}^{(-)}(pw_{0},pw)H_{1}(q,\alpha
_{0},2rp),  \label{FRbJm}
\end{equation}%
with the function $H_{1}(q,\alpha _{0},x)$ from (\ref{Hnq}). Here, the
functions in the integrand are defined as%
\begin{eqnarray}
F_{\mathrm{(R)}}^{(-)}(x,y) &=&\frac{I_{\nu _{1}}(x)}{K_{\nu _{1}}(x)}K_{\nu
_{1}}(y)K_{\nu _{2}}(y),  \notag \\
F_{\mathrm{(L)}}^{(-)}(x,y) &=&\frac{K_{\nu _{2}}(x)}{I_{\nu _{2}}(x)}I_{\nu
_{1}}(y)I_{\nu _{2}}(y),  \label{FRLm}
\end{eqnarray}%
with $\nu _{1}$ and $\nu _{2}$ from (\ref{n12}). Comparing with (\ref{FRL})
and (\ref{FCbJ2}), we see that, for a given $s$, the brane-induced
contribution for the field obeying the equation (\ref{DiracEq}) and the
condition (\ref{BCm}) differs from the FC for the field, obeying the field
equation (\ref{DiracEq}) with $s$ replaced by $-s$ and the condition (\ref%
{MITbc}), only in the sign. The same property is the case for the terms $%
\langle \bar{\psi}\psi \rangle ^{\mathrm{AdS}}$ and $\langle \bar{\psi}\psi
\rangle _{\mathrm{cs}}$: they change the signs under the replacement $%
s\rightarrow -s$.

In a similar way we can see that the VEV\ of the energy-momentum tensor for
the field obeying the equation (\ref{DiracEq}) and the boundary condition (%
\ref{BCm}) is presented as (\ref{Tmudec}) with the brane-induced
contribution (no summation over $\mu $)%
\begin{equation}
\langle T_{\mu }^{\mu }\rangle _{\mathrm{b,J}}=\frac{w^{6}}{\pi ^{2}a^{5}}%
\int_{0}^{\infty }dp\,p^{5}F_{\mathrm{(J)}}^{(-)(\mu )}(pw_{0},pw)H^{(\mu
)}(q,\alpha _{0},2pr),  \label{TmuJ5m}
\end{equation}%
where the functions $H^{(\mu )}(q,\alpha _{0},x)$ are given by (\ref{Hmu}).
In the R-region the functions in the integrand are defined as%
\begin{eqnarray}
F_{\mathrm{(R)}}^{(-)(\mu )}(x,y) &=&\frac{I_{\nu _{1}}(x)}{K_{\nu _{1}}(x)}%
\left[ K_{\nu _{2}}^{2}(y)-K_{\nu _{1}}^{2}(y)\right] ,\;\mu =0,1,2,4,
\notag \\
F_{\mathrm{(R)}}^{(-)(3)}(x,y) &=&\frac{I_{\nu _{1}}(x)}{K_{\nu _{1}}(x)}%
\left[ K_{\nu _{1}}^{2}(y)-K_{\nu _{2}}^{2}(y)+\frac{2sma}{y}K_{\nu
_{1}}(y)K_{\nu _{2}}(y)\right] .  \label{FR3m}
\end{eqnarray}%
and for the L-region%
\begin{eqnarray}
F_{\mathrm{(L)}}^{(-)(\mu )}(x,y) &=&\frac{K_{\nu _{2}}(x)}{I_{\nu _{2}}(x)}%
\left[ I_{\nu _{1}}^{2}(y)-I_{\nu _{2}}^{2}(y)\right] ,\;\mu =0,1,2,4,
\notag \\
F_{\mathrm{(L)}}^{(-)(3)}(x,y) &=&\frac{K_{\nu _{2}}(x)}{I_{\nu _{2}}(x)}%
\left[ I_{\nu _{2}}^{2}(y)-I_{\nu _{1}}^{2}(y)+\frac{2sma}{y}I_{\nu
_{1}}(y)I_{\nu _{2}}(y)\right] .  \label{FL3m}
\end{eqnarray}%
Comparison of this result with (\ref{F3R}), (\ref{F3L}) and (\ref{TmuJ5})
shows that the brane-induced VEV of the energy-momentum tensor for the field
with a given $s$ and obeying the boundary condition (\ref{BCm}) coincides
with the corresponding quantity for the field with $s$ replaced by $-s$ in
the field equation (\ref{DiracEq}) and obeying the condition (\ref{MITbc}).
We recall that the brane-free contribution in the VEV of the energy-momentum
tensor does not depend on $s$.

Equivalently, the obtained features can be formulated as follows: The FC for
the field obeying the field equation (\ref{DiracEq}) and the boundary
condition $(1+si\gamma ^{\mu }n_{\mu })\psi =0$ on the brane $w=w_{0}$ is an
odd function of $s$, whereas the VEV of the energy-momentum tensor is an
even function.

\subsection{VEVs for the second representation and applications in RSII
braneworld}

For the representation of the Clifford algebra corresponding to $s=-1$ the
transformed field $\psi _{(-1)}^{\prime }$ obeys the field equation (\ref%
{DiracEq}) with $s=-1$ and the boundary condition (\ref{BCm}). Hence, in
accordance with the result formulated in the previous subsection, we
conclude that the corresponding FC is expressed as
\begin{equation}
\langle \bar{\psi}_{(-1)}^{\prime }\psi _{(-1)}^{\prime }\rangle =-\langle
\bar{\psi}_{(+1)}\psi _{(+1)}\rangle =-\langle \bar{\psi}\psi \rangle
_{s=+1},  \label{FCsec}
\end{equation}%
where $\langle \bar{\psi}\psi \rangle _{s=+1}$ is given by the expressions
in Section \ref{sec:FCond} with $s=+1$. In order to find the FC for the
initial field $\psi _{(-1)}$ we make the inverse transformation of the
field. It is easy to see that $\langle \bar{\psi}_{(-1)}\psi _{(-1)}\rangle
=-\langle \bar{\psi}_{(-1)}^{\prime }\psi _{(-1)}^{\prime }\rangle $.
Combining with (\ref{FCsec}) we conclude that the FCs coincide for the
fields realizing two inequivalent irreducible representations of the
Clifford algebra if they obey the boundary condition (\ref{BCs}): $\langle
\bar{\psi}_{(+1)}\psi _{(+1)}\rangle =\langle \bar{\psi}_{(-1)}\psi
_{(-1)}\rangle $. The corresponding expressions are given by the formulas in
Section \ref{sec:FCond} with $s=+1$. The same is the case for the fields $%
\psi _{(s)}$ obeying the boundary condition
\begin{equation}
(1-i\gamma _{(s)}^{\mu }n_{\mu })\psi _{(s)}=0  \label{BCsm}
\end{equation}%
for $w=w_{0}$. The corresponding FC is given by the formula (\ref{FRbJm})
with $s=+1$ (or by the formulas in Section \ref{sec:FCond} with $s=-1$ and
with the change of the sign in front of all the terms). In a similar way, we
can see that the VEVs of the energy-momentum tensor for the fields $\psi
_{(\pm 1)}$ obeying the boundary condition (\ref{BCs}) coincide with the
VEVs for the fields $\psi _{(\pm 1)}^{\prime }$ obeying the condition (\ref%
{BCsp}). They are given by the formulas in Section \ref{sec:EMT0} with $s=+1$
for the field $\psi _{(s)}$ obeying the boundary condition (\ref{BCs}) and
by the formula (\ref{TmuJ5m}) with $s=+1$ (or equivalently by (\ref{TmuJ5})
with $s=-1$) for the boundary condition (\ref{BCsm}).

The mass term in the Lagrangian density $L_{(s)}$ is not invariant under the
charge conjugation ($C$) and the parity transformation ($P$). In the absence
of an external gauge field, we can combine two fields $\psi _{(+1)}$ and $%
\psi _{(-1)}$ for the construction of fermionic models with the Lagrangian
density $L=\sum_{s=\pm 1}L_{(s)}$ invariant under those transformations. The
total FC and the VEV of the energy-momentum tensor are obtained by summing
the corresponding VEVs for separate fields. They are given by the formulas
in Sections \ref{sec:FCond} and \ref{sec:EMT0}, with an additional
coefficient 2, for $s=+1$ in the case of the boundary conditions (\ref{BCs})
and for $s=-1$ in the case of conditions (\ref{BCsm}) (with an additional
change in the sign of the FC for (\ref{BCsm})).

As a realization of the model under consideration we can consider the
Randall-Sundrum model with a single brane (RSII model) \cite{Rand99b} in the
presence of a topological defect. The brane in the corresponding setup is
located at $y=0$ and the background geometry contains two copies of the
R-region identified by the $Z_{2}$-symmetry. The line element is obtained
from (\ref{ds1}) by the replacement $e^{-2y/a}\rightarrow e^{-2|y|/a}$. The
fields in the regions $-\infty <y<0$ and $0<y<+\infty $ are related by the $%
Z_{2}$-symmetry of the model. The $4\times 4$ matrix $M$ in the relation $%
\psi (t,r,\phi ,-y,z)=M\psi (t,r,\phi ,y,z)$ is determined by the
requirement of the invariance of the action under the $Z_{2}$ identification
(see \cite{Bell18,Flac01b}). The following conditions are obtained on the
matrix $M$: $\{\gamma ^{(0)},M\}=0$, $\{\gamma ^{(0)}\gamma ^{(3)},M\}=0$,
and $[\gamma ^{(0)}\gamma ^{(b)},M]=0$ with $b=1,2,4$. The corresponding
solution with an additional constraint $M^{2}=1$ is given as $M=\pm \mathrm{%
diag}(\sigma ^{3},-\sigma ^{3})=\pm i\gamma ^{(3)}$. Two types of fermionic
fields in braneworld models correspond to the upper and lower signs in the
expression for the matrix $M$. The boundary conditions on the brane for
those fields are obtained by taking $y=0$ in the relation mapping two copies
of the R-region. It is reduced to $\left( 1\mp i\gamma ^{(3)}\right) \psi
(x)=0$. By taking into account that in the R-region we had $n_{\mu }=-\delta
_{\mu }^{3}a/w$, we see that for the field with $M=i\gamma ^{(3)}$ the
boundary condition dictated by the $Z_{2}$-symmetry is reduced to (\ref%
{MITbc}). For the field corresponding to $M=-i\gamma ^{(3)}$ we get the
boundary condition (\ref{BCm}). The FC and the VEV of the energy-momentum
tensor in the RSII model with a topological defect (for quantum effects in
higher-dimensional generalizations of RSII model see \cite{Saha20} and
references therein) are expressed by the formulas given above for respective
boundary conditions taking $w_{0}=a$ and with an additional coefficient 1/2.
The latter is related to the fact that in $Z_{2}$-symmetric braneworld model
the integral over $y$ in the normalization condition for the fermionic modes
goes over the region $-\infty <y<+\infty $ instead of the region $y\in
\lbrack y_{0}=0,\infty )$ in our consideration above for the R-region.

In the context of the braneworld scenario, for an observer located on the
brane at $y=0$ the induced line element takes the form $ds_{\mathrm{b}%
}^{2}=dt^{2}-dr^{2}-r^{2}d\phi ^{2}-dz^{2}$ with $0\leq \phi \leq 2\pi /q$.
This is the line element for the geometry corresponding to a straight cosmic
string in (3+1)-dimensional flat spacetime. From the point of view of
physics on the brane it is of interest to compare the VEVs induced on the
brane by the topological defect in the background AdS spacetime with the
corresponding VEVs\ induced by a cosmic string on the Minkowski bulk with
the line element $ds_{\mathrm{b}}^{2}$. As it has been discussed above, the
VEVs $\langle \bar{\psi}\psi \rangle _{\mathrm{J}}$ and $\left\langle T_{\mu
}^{\mu }\right\rangle _{\mathrm{J}}$ diverge on the brane. The divergences
come from the purely brane-induced parts $\langle \bar{\psi}\psi \rangle _{%
\mathrm{J}}^{(0)}$ and $\left\langle T_{\mu }^{\mu }\right\rangle _{\mathrm{J%
}}^{(0)}$. They are absorbed by the renormalization of the on-brane FC and
energy-momentum tensor in the absence of the cosmic string. The
corresponding renormalized VEVs do not depend on the characteristics of the
cosmic string. The contributions in the VEVs induced by the cosmic string on
the AdS bulk, given by $\langle \bar{\psi}\psi \rangle _{\mathrm{J}}-\langle
\bar{\psi}\psi \rangle _{\mathrm{J}}^{(0)}$ and $\langle T_{\mu }^{\mu
}\rangle _{\mathrm{J}}-\langle T_{\mu }^{\mu }\rangle _{\mathrm{J}}^{(0)}$,
are finite on the brane. Moreover, as it has been shown above, that
contribution in the FC vanishes on the brane. For the components with $\mu
=0,3$, the cosmic string-induced contribution $\langle T_{\mu }^{\mu
}\rangle _{\mathrm{J}}-\langle T_{\mu }^{\mu }\rangle _{\mathrm{J}}^{(0)}$
on the brane $y=0$ is directly found from the representation (\ref{TmuJ3})
taking $w=w_{0}=a$ (with a coefficient 1/2 for RSII model). For these values
and for the R-region one has $W_{\nu _{1}}^{(0)}(pa)=4/\left( \pi pa\right)
^{2}$ and $W_{\nu _{1}}^{(3)}(pa)=-4/\pi ^{2}a^{2}$. The other components
are found by using the relations (\ref{T014}) and (\ref{T2}). The
corresponding expressions are further simplified for a massless field. In
that case $J_{\nu _{2}}^{2}(pa)+Y_{\nu _{2}}^{2}(pa)=2/(\pi pa)$ and the
integral over $p$ in (\ref{TmuJ3}) for the R-region is evaluated by using
the formula $\int_{0}^{\infty }dx\,x^{2+\delta _{\mu 3}}K_{2-\delta _{\mu
3}}(x)=3\pi /2$. For the cosmic string-induced part on the brane this gives
(no summation over $\mu $)
\begin{equation}
\left[ \langle T_{\mu }^{\mu }\rangle _{\mathrm{R}}-\langle T_{\mu }^{\mu
}\rangle _{\mathrm{R}}^{(0)}\right] _{w=a}=\frac{3h_{5}(q,\alpha _{0})}{%
32\pi ^{2}r^{5}}\mathrm{diag}(1,1,-4,1,1).  \label{Tmubr}
\end{equation}%
Note that this result for a massless field does not depend on the curvature
radius of the AdS spacetime. For a massive field, the curvature radius
appears in the expression for the product $a^{5}\left[ \langle T_{\mu }^{\mu
}\rangle _{\mathrm{R}}-\langle T_{\mu }^{\mu }\rangle _{\mathrm{R}}^{(0)}%
\right] _{w=a}$ in the form of dimensionless combinations $ma$ and $r/a$.
The expectation values induced on the brane by quantum fluctuations of bulk
fields differ from those for a cosmic string in Minkowski spacetime with the
line element $ds_{\mathrm{b}}^{2}$. The VEVs for the latter geometry in the
absence of magnetic flux have been considered in \cite{Beze08cyl}. The
corresponding FC is nonzero for a massive fermionic field and vanishes in
the massless limit. The VEV of the energy-momentum tensor is different from
zero for both massless and massive fields. For a massless field it is
reduced to the result found in \cite{Frol87}.

\section{Conclusion}

\label{Conc}

We have investigated the combined effects of a cosmic string and of a brane
parallel to the AdS boundary on the local properties of the fermionic
vacuum. As representatives of those properties the FC and the VEV of the
energy-momentum tensor are considered. For the evaluation of the
corresponding expectation values the direct summation over the complete set
of the fermionic modes from \cite{Bell21} has been used. They are specified
by the set of quantum numbers $(\lambda ,j,p,k_{z},\eta )$. The eigenvalues
of the quantum number $p$ are continuous in the R-region and discrete in the
L-region and the properties of the vacuum are different in those regions.
Both the FC and the vacuum energy-momentum tensor are decomposed into three
separate contributions. The first one corresponds to the pure AdS geometry
when the cosmic string and the brane are absent and due to the maximal
symmetry of the AdS spacetime and of the corresponding vacuum state it does
not depend on a spacetime point. The second contribution to the VEVs
presents the part that is induced by the cosmic string in the brane-free
geometry. Those contribution for the FC and the energy-momentum tensor have
been investigated in \cite{Bell21b} and \cite{Bell22}, respectively. Our
main interest in the present paper is concentrated on the contributions in
the FC and the energy-momentum tensor induced by the presence of the brane.
In order to explicitly separate those contributions in the L-region we have
used the generalized Abel-Plana formula for the summation of series over the
zeros of the Bessel functions (related to the eigenvalues of the quantum
number $p$). The corresponding representation for the R-region is obtained
by an appropriate rotation of the integration contour in the integral over
continuous eigenvalues of $p$. After those transformations the brane-induced
contributions are given by (\ref{FCbJ2}) for the FC and by (\ref{TmuJ5}) for
the energy-momentum tensor. They are even periodic functions of the magnetic
flux with the period equal to the flux quantum.

In order to clarify the behavior of the brane-induced contributions in the
VEVs we have considered limiting cases and asymptotic regions of the
parameters. The general formulas are simplified in two special cases. The
first one corresponds to the absence of the magnetic flux with $\alpha
_{0}=0 $ and the second one corresponds to the absence of the planar angle
deficit with $q=1$. The respective expressions for the FC and the VEV of the
energy-momentum tensor are obtained from the expressions (\ref{FCbJ2}) and (%
\ref{TmuJ5}) substituting the functions (\ref{Hnq}) from (\ref{Hnqalf0}) and
(\ref{Hnq1}), respectively. In the limit $a\rightarrow \infty $, with fixed
values for $y$ and $y_{0}$, we have obtained the VEVs in the geometry of a
cosmic string in background of (4+1)-dimensional Minkowski spacetime in the
presence of a planar boundary on which the field obeys the bag boundary
condition. For a massless fermionic field the problem under consideration is
conformally related to the corresponding problem in the Minkowski bulk with
a single boundary for the R-region and with two parallel boundaries in the
L-region. The one of the boundaries in the latter case is the conformal
image of the brane and the second boundary is the conformal image of the AdS
boundary.

The brane-induced contributions in the VEVs are mainly located in the region
near the brane. For the FC the leading term in the corresponding asymptotic
expansion is given by (\ref{FCnearBr}). Near the brane the effects of the
background curvature and of the mass are weak and the leading term (\ref%
{FCnearBr}) coincides with that for a boundary in the Minkowski bulk in the
absence of cosmic string. For the R-region, the large values of the ratio $%
w/w_{0}$ correspond to large proper distances from the brane compared with
the curvature radius of the background spacetime. For a given value of the
ratio $r/w$, the brane-induced VEVs for the FC and the energy-momentum
tensor behave like $(w_{0}/w)^{\nu _{2}+|\nu _{2}|}$. In particular, this
shows that when the location of the brane tends to the AdS boundary the
brane-induced contributions vanish as $w_{0}^{1+2ma}$ for the field with $%
s=+1$. For the field with $s=-1$ that contribution vanishes like $%
w_{0}^{2ma-1}$ in the range of the mass $ma>1/2$ and tends to a nonzero
finite value for $ma<1/2$. In the L-region, the large proper distances from
the brane correspond to small values of the ratio $w/w_{0}\gg 1$ and the
brane-induced parts decay as $(w/w_{0})^{5+2ma}$.

Near the cosmic string and for massive fields the VEVs are dominated by the
contributions $\langle \bar{\psi}\psi \rangle _{\mathrm{cs}}$ and $\langle
T_{\mu }^{\nu }\rangle _{\mathrm{cs}}$. They behave like $(r/a)^{3}$ and $%
(r/a)^{5}$ for the FC and energy-momentum tensor, respectively. For the
energy-momentum tensor the leading behavior remains the same for a massless
field, whereas for the FC the leading term vanishes and near the cosmic
string the contributions $\langle \bar{\psi}\psi \rangle _{\mathrm{cs}}$ and
$\langle \bar{\psi}\psi \rangle _{\mathrm{b,J}}$ are of the same order. The
brane-induced contributions $\langle \bar{\psi}\psi \rangle _{\mathrm{b,J}}$
and $\langle T_{\mu }^{\mu }\rangle _{\mathrm{b,J}}$ vanish on the cosmic
string for $2|\alpha _{0}|<1-1/q$, take a finite limiting value for $%
2|\alpha _{0}|=1-1/q$ and diverge as $\left( r/w\right) ^{(1-2|\alpha
_{0}|)q-1}$ in the range $2|\alpha _{0}|>1-1/q$. The effects induced by the
cosmic string in the FC and in the VEV\ of the energy-momentum tensor are
described by the differences $\langle \bar{\psi}\psi \rangle _{\mathrm{J}%
}-\langle \bar{\psi}\psi \rangle _{\mathrm{J}}^{(0)}$ and $\left\langle
T_{\mu }^{\mu }\right\rangle _{\mathrm{J}}-\left\langle T_{\mu }^{\mu
}\right\rangle _{\mathrm{J}}^{(0)}$. The behavior of those quantities at
large distances from the string essentially differs in the R- and L-regions.
For the R-region and in the cases $s=+1$ or $s=-1$, $ma<1/2$, the cosmic
string-induced effects decay as $(r/w)^{5+2sma}$, whereas for the case $s=-1$%
, $ma>1/2$ the decay is slower, like $(r/w)^{3+2ma}$. It is of interest to
note that, considered as a function of the proper distance from the string,
the decay of the cosmic string-induced contributions in the VEVs at large
distances follows a power-law for both the cases of massless and massive
fields. This behavior is in contrast to the one for the Minkowski bulk with
the exponential decay for massive fields. The eigenvalues of the quantum
number $p$ in the L-region are discrete and the cosmic string-induced VEVs
at large distances are suppressed by the exponential factor $e^{-2\left(
r/w_{0}\right) p_{1}}$, where $p_{1}/w_{0}$ is the lowest eigenvalue for $p$.

We have considered an odd dimensional background spacetime and the
corresponding Clifford algebra for the gamma matrices has two inequivalent
irreducible representations. The FC and the VEV\ of the energy-momentum
tensor for the fields realizing those representations, denoted here as $\psi
_{(+1)}$ and $\psi _{(-1)}$, are obtained from the results we have provided
in Sections \ref{sec:FCond} and \ref{sec:EMT0}. If the bag boundary
condition is imposed on the brane for both the fields (see (\ref{BCs})) and
they have equal massess, then the VEVs for the fields $\psi _{(+1)}$ and $%
\psi _{(-1)}$coincide and they are given by the formulas in Sections \ref%
{sec:FCond} and \ref{sec:EMT0} for $s=+1$. If the fields $\psi _{(+1)}$ and $%
\psi _{(-1)}$ obey the boundary condition (\ref{BCsm}), that differs from
the bag boundary condition by the sign of the term containing the normal to
the brane, then the corresponding VEVs for both the fields are obtained from
(\ref{FRbJm}) and (\ref{TmuJ5m}) with $s=+1$. Equivalently, the VEVs are
obtained from the formulae in Sections \ref{sec:FCond} and \ref{sec:EMT0}
with $s=-1$ additionally changing the sign for the FC. We have seen that two
types of the considered boundary conditions naturally arise in
Randall-Sundrum type braneworld models as a consequence of the $Z_{2}$%
-symmetry with respect to the brane. The FC and the VEV of the
energy-momentum tensor in RSII model in the presence of cosmic string are
obtained directly from the formulas given above.

\section*{Acknowledgments}

A.A.S. was supported by the grants No. 20RF-059 and No. 21AG-1C047 of the
Science Committee of the Ministry of Education, Science, Culture and Sport
RA. E.R.B.M. is partially supported by CNPQ under Grant no. 301.783/2019-3.

\appendix

\section{Extraction of the brane-induced contributions}

\label{sec:App1}

\subsection{Fermion condensate}

In order to extract the parts in the FC induced by the brane we need to
evaluate the difference%
\begin{equation}
f_{\mathrm{(J)}}(w_{0},w,\gamma )=s\sum_{(p)}p^{2}\frac{W_{\nu
_{1}}(pw)W_{\nu _{2}}(pw)}{w_{0}^{2}U_{\nu _{2}}^{\mathrm{(J)}}(pw_{0})}%
K_{1}(p\gamma )-s\int_{0}^{\infty }dp\,p^{3}J_{\nu _{1}}(pw)J_{\nu
_{2}}(pw)K_{1}(p\gamma ).  \label{fJ}
\end{equation}%
First let us consider the R-region. By taking into account that $W_{\nu
}(pw)=G_{\nu _{2},\nu }(pw_{0},pw)$ and using (\ref{UJ}), (\ref{Sump}), the
following relation can be checked:
\begin{equation}
f_{\mathrm{(R)}}(w_{0},w,\gamma )=-\frac{s}{2}\sum_{n=1,2}\int_{0}^{\infty
}dp\,p^{3}\frac{J_{\nu _{2}}(pw_{0})}{H_{\nu _{2}}^{(n)}(pw_{0})}H_{\nu
_{1}}^{(n)}(pw)H_{\nu _{2}}^{(n)}(pw)K_{1}(p\gamma )\ ,  \label{fR}
\end{equation}%
with $H^{(n)}(z)$, $n=1,2$, being the Hankel functions \cite{Abra}. As the
next step, we rotate the integration contour in (\ref{fR}) by the angle $\pi
/2$ for the term with $n=1$ and by the angle $-\pi /2$ for $n=2$.
Introducing the modified Bessel functions we get%
\begin{equation}
f_{\mathrm{(R)}}(w_{0},w,\gamma )=\int_{0}^{\infty }dp\,p^{3}\frac{I_{\nu
_{2}}(pw_{0})}{K_{\nu _{2}}(pw_{0})}K_{\nu _{1}}(pw)K_{\nu
_{2}}(pw)J_{1}(p\gamma )\ .  \label{fR2}
\end{equation}

For the L-region the functions in (\ref{fJ}) are given by (\ref{Wn}), (\ref%
{UJ}), and $\sum_{(p)}=\sum_{i}$, $p=p_{i}/w_{0}$. In this case we use a
variant of the generalized Abel-Plana formula \cite{Saha07}:
\begin{equation}
\sum_{i=1}^{\infty }\frac{2f(p_{i})}{p_{i}J_{\nu _{2}}^{2}(p_{i})}%
=\int_{0}^{\infty }duf(u)-\frac{1}{\pi }\int_{0}^{\infty }du\,\frac{K_{\nu
_{1}}(u)}{I_{\nu _{1}}(u)}[e^{-\nu _{1}\pi i}f(e^{\pi i/2}u)+e^{\nu _{1}\pi
i}f(e^{-\pi i/2}u)]\ ,  \label{APF}
\end{equation}%
valid for a function $f(u)$ analytic in the right half-plane of the complex
variable $u$ (additional conditions imposed on the function $f(u)$ can be
found in \cite{Saha07}). For the evaluation of the difference (\ref{fJ}) we
take
\begin{equation}
f(u)=u^{3}J_{\nu _{1}}(uw/w_{0})J_{\nu _{2}}(uw/w_{0})K_{1}(u\gamma /w_{0})\
.  \label{fL}
\end{equation}%
This gives%
\begin{equation}
f_{\mathrm{(L)}}(w_{0},w,\gamma )=\int_{0}^{\infty }dp\,p^{3}\frac{K_{\nu
_{1}}(pw_{0})}{I_{\nu _{1}}(pw_{0})}I_{\nu _{1}}(pw)I_{\nu
_{2}}(pw)J_{1}(p\gamma ).  \label{fL2n}
\end{equation}

\subsection{Energy-momentum tensor}

For the energy-momentum tensor the functions in (\ref{TmuJ4}) are defined as
\begin{equation}
f_{\mathrm{(J)}}^{(\mu )}(w_{0},w,\gamma )=\sum_{(p)}p^{2-\delta _{3\mu }}%
\frac{W_{\nu _{1}}^{(\mu )}(pw)}{w_{0}^{2}U_{\nu _{2}}^{\mathrm{(J)}}(pw_{0})%
}K_{2-\delta _{3\mu }}(p\gamma )-\int_{0}^{\infty }dp\,p^{3-\delta _{3\mu
}}W_{0,\nu _{1}}^{(\mu )}(pw)K_{2-\delta _{3\mu }}(p\gamma ).  \label{fJmu}
\end{equation}%
for $\mu =0,3$. In the R-region they can be presented in the form%
\begin{eqnarray}
f_{\mathrm{(R)}}^{(0)}(w_{0},w,\gamma ) &=&-\frac{1}{2}\sum_{n=1,2}%
\int_{0}^{\infty }dp\,p^{2}\frac{J_{\nu _{2}}(pw_{0})}{H_{\nu
_{2}}^{(n)}(pw_{0})}K_{2}(p\gamma )\left[ H_{\nu _{1}}^{(n)2}(pw)+H_{\nu
_{2}}^{(n)2}(pw)\right] ,  \notag \\
f_{\mathrm{(R)}}^{(3)}(w_{0},w,\gamma ) &=&\frac{1}{2}\sum_{n=1,2}\int_{0}^{%
\infty }dp\,p^{4}\frac{J_{\nu _{2}}(pw_{0})}{H_{\nu _{2}}^{(n)}(pw_{0})}%
K_{1}(p\gamma )  \notag \\
&&\times \left[ H_{\nu _{1}}^{(n)2}(pw)+H_{\nu _{2}}^{(n)2}(pw)-\frac{2ma}{pw%
}H_{\nu _{1}}^{(n)}(pw)H_{\nu _{2}}^{(n)}(pw)\right] .  \label{fR3}
\end{eqnarray}%
After the rotation of the integration contour by the angles $\pi /2$ and $%
-\pi /2$ for the parts with $n=1$ and $n=2$, respectively, the functions are
transformed as%
\begin{equation}
f_{\mathrm{(R)}}^{(\mu )}(w_{0},w,\gamma )=\left( -1\right) ^{\delta _{0\mu
}}\int_{0}^{\infty }dp\,p^{3+\delta _{3\mu }}F_{\mathrm{(R)}}^{(\mu
)}(pw_{0},pw)J_{2-\delta _{3\mu }}(p\gamma ),  \label{fRmu}
\end{equation}%
for $\mu =0,3$. Here, the functions $F_{\mathrm{(R)}}^{(\mu )}(pw_{0},pw)$
are given by (\ref{F3R}).

In the L-region $\sum_{(p)}=\sum_{i=1}^{\infty }$ and for the summation of
the series we use the Abel-Plana formula (\ref{APF}). Introducing the
modified Bessel functions one gets%
\begin{equation}
f_{\mathrm{(L)}}^{(\mu )}(w_{0},w,\gamma )=\left( -1\right) ^{\delta _{0\mu
}}\int_{0}^{\infty }dp\,p^{3+\delta _{3\mu }}F_{\mathrm{(L)}}^{(\mu
)}(pw_{0},pw)J_{2-\delta _{3\mu }}(p\gamma ),  \label{fLmu}
\end{equation}%
where $\mu =0,3$ and the functions $F_{\mathrm{(L)}}^{(\mu )}(x,y)$ are
defined by (\ref{F3L}).


\begin{thebibliography}{99}
\bibitem{Eliz94C} E. Elizalde, S. D. Odintsov, A. Romeo, A. A. Bytsenko, and
S. Zerbini, \textit{Zeta Regularization Techniques with Applications} (World
Scientific, Singapore, 1994); V. M. Mostepanenko and N. N. Trunov, \textit{%
The Casimir Effect and Its Applications} (Clarendon, Oxford, 1997); K. A.
Milton, \textit{The Casimir Effect: Physical Manifestation of Zero-Point
Energy} (World Scientific, Singapore, 2002); M. Bordag, G. L. Klimchitskaya,
U. Mohideen, and V. M. Mostepanenko, \textit{Advances in the Casimir Effect}
(Oxford University Press, New York, 2009); \textit{Casimir Physics}, edited
by D. Dalvit, P. Milonni, D. Roberts, and F. da Rosa, Lecture Notes in
Physics Vol. 834 (Springer-Verlag, Berlin, 2011).

\bibitem{Maar10} R. Maartens and K. Koyama, Brane-world gravity, Living
Reviews in Relativity \textbf{13}, 5 (2010).

\bibitem{Ahar00} O. Aharony, S. S. Gubser, J. Maldacena, H. Ooguri, and Y.
Oz, Large N field theories, string theory and gravity, Phys. Rep. 323, 183
(2000); H. N\u{a}stase, \textit{Introduction to AdS/CFT Correspondence}
(Cambridge University Press, Cambridge, England, 2015); M. Ammon and J.
Erdmenger, \textit{Gauge/Gravity Duality: Foundations and Applications}
(Cambridge University Press, Cambridge, England, 2015).

\bibitem{Vile94} A.Vilenkin, E.P.S. Shellard, \textit{Cosmic Strings and
Other Topological Defects} (Cambridge University Press, Cambridge, 1994); M.
B. Hindmarsh and T. W. B. Kibble, Cosmic strings, Rep. Prog. Phys. \textbf{58%
}, 477 (1995).

\bibitem{Witt85} E. Witten, Cosmic superstrings, Phys. Lett. B \textbf{153},
243 (1985); G. Dvali and S. H. Henry Tye, Brane inflation, Phys. Lett. B%
\textbf{\ 450}, 72 (1999); S. Sarangi and S. H. Henry Tye, Cosmic string
production towards the end of brane inflation, Phys. Lett. B\textbf{\ 536},
185 (2002); E. J. Copeland and T. W. B. Kibble, Cosmic strings and
superstrings, Proc. R. Soc. A \textbf{466}, 623 (2010); E. J. Copeland, L.
Pogosian, and T.Vachaspati, Seeking string theory in the cosmos, Class.
Quantum Grav. \textbf{28}, 204009 (2011); D. F. Chernoff and S. H. Henry
Tye, Inflation, string theory and cosmic strings, Int. J. Mod. Phys. D
\textbf{24}, 1530010 (2015).

\bibitem{Dehg01} M. H. Dehghani, A. M. Ghezelbash, and R. B. Mann, Vortex
holography, Nucl. Phys. B \textbf{625}, 389 (2002); C. A. Ballon Bayona, C.
N. Ferreira, and V. J. Vasquez Otoya, A conical deficit in the AdS$_{4}$/CFT$%
_{3}$ correspondence, Class. Quantum Grav. \textbf{28}, 015011 (2011); A. de
P\'{a}dua Santos and E. R. Bezerra de Mello, Non-Abelian cosmic strings in
de Sitter and anti-de Sitter space, Phys. Rev. D \textbf{94}, 063524 (2016).

\bibitem{Davi01} S. C. Davis, Brane world linearized cosmic string gravity,
Phys. Lett. B \textbf{499}, 179 (2001); S. C. Davis, Brane world cosmic
string interaction, Phys. Lett. B \textbf{645}, 323 (2007); M. Heydari-Fard,
H. Razmi, and S. Y. Rokni, Brane-world cosmic strings revive the
cosmological constant, Class. Quantum Grav. \textbf{30}, 165001 (2013); M.
C. B. Abdalla, M. E. X. Guimar\~{a}es, and J. M. H. da Silva, Brane cosmic
string compactification in Brans-Dicke theory, Phys. Rev. D \textbf{75},
084028 (2007); M. C. B. Abdalla, P. F. Carlesso, and J. M. Hoff da Silva,
Solution for a local straight cosmic string in the braneworld gravity, Eur.
Phys. J. C \textbf{75}, 432 (2015).

\bibitem{Bell14} S. Bellucci, E. R. Bezerra de Mello, A. de Padua, and A. A.
Saharian, Fermionic vacuum polarization in compactified cosmic string
spacetime, Eur. Phys. J. C 74, 2688 (2014).

\bibitem{Mota18} H. F. Mota, E. R. Bezerra de Mello, and K. Bakke, Scalar
Casimir effect in a high-dimensional cosmic dispiration spacetime, Int. J.
Mod. Phys. D \textbf{27}, 1850107 (2018).

\bibitem{Beze09} E. R. Bezerra de Mello and A. A. Saharian, Vacuum
polarization by a cosmic string in de Sitter spacetime, J. High Energy Phys.
04 (2009) 046; E. R. Bezerra de Mello and A. A. Saharian, Fermionic vacuum
polarization by a cosmic string in de Sitter spacetime, J. High Energy Phys.
08 (2010) 038; A. Mohammadi, E. R. Bezerra de Mello, and A.A. Saharian,
Induced fermionic currents in de Sitter spacetime in the presence of a
compactified cosmic string, Class. Quantum Gravity \textbf{32}, 135002
(2015); A. A. Saharian, V. F. Manukyan, and N. A. Saharyan, Electromagnetic
vacuum fluctuations around a cosmic string in de Sitter spacetime, Eur.
Phys. J. C \textbf{77}, 478 (2017).

\bibitem{Beze12} E. R. Bezerra de Mello and A.A. Saharian, Vacuum
polarization induced by a cosmic string in anti-de Sitter spacetime, J.
Phys. A \textbf{45}, 115402 (2012); E. R. Bezerra de Mello, E. R. Figueiredo
Medeiros, and A. A. Saharian, Fermionic vacuum polarization by a cosmic
string in Anti-de Sitter spacetime, Class. Quant. Grav. \textbf{30}, 175001
(2013).

\bibitem{Bell21} S. Bellucci, W. Oliveira dos Santos, E.R. Bezerra de Mello,
and A.A. Saharian, Vacuum fermionic currents in braneworld models on AdS
bulk with a cosmic string, J. High Energy Phys. 02 (2021) 190.

\bibitem{Flac03} A. Flachi, J. Garriga, O. Pujol\`{a}s, and T. Tanaka,
Moduli stabilization in higher dimensional brane models, J. High Energy
Phys. 08 (2003) 053; A. Flachi and O. Pujol\`{a}s, Quantum self-consistency
of AdS $\times \Sigma $ brane models, Phys. Rev. D \textbf{68}, 025023
(2003); A. A. Saharian, Wightman function and vacuum fluctuations in higher
dimensional brane models, Phys. Rev. D \textbf{73}, 044012 (2006); A. A.
Saharian, Bulk Casimir densities and vacuum interaction forces in higher
dimensional brane models, Phys. Rev. D \textbf{73}, 064019 (2006); A. A.
Saharian, Surface Casimir densities and induced cosmological constant in
higher dimensional braneworlds, Phys. Rev. D \textbf{74}, 124009 (2006); E.
Elizalde, M. Minamitsuji, and W. Naylor, Casimir effect in rugby-ball type
flux compactifications, Phys. Rev. D 75, 064032 (2007); R. Linares, H. A.
Morales-T\'{e}cotl, and O. Pedraza, Casimir force for a scalar field in
warped brane worlds, Phys. Rev. D \textbf{77}, 066012 (2008); M. Frank, N.
Saad, and I. Turan, Casimir force in Randall-Sundrum models with q+1
dimensions, Phys. Rev. D \textbf{78}, 055014 (2008).

\bibitem{Beze15} E. R. Bezerra de Mello, A. A. Saharian, and V. Vardanyan,
Induced vacuum currents in anti-de Sitter space with toral dimensions, Phys.
Lett. B \textbf{741}, 155 (2015); S. Bellucci, A. A. Saharian, and V.
Vardanyan, Vacuum currents in braneworlds on AdS bulk with compact
dimensions, J. High Energy Phys. 11 (2015) 092; S. Bellucci, A. A. Saharian,
and V. Vardanyan, Hadamard function and the vacuum currents in braneworlds
with compact dimensions: Two-brane geometry, Phys. Rev. D \textbf{93},
084011 (2016); S. Bellucci, A. A. Saharian, and V. Vardanyan, Fermionic
currents in AdS spacetime with compact dimensions, Phys. Rev. D \textbf{96},
065025 (2017); S. Bellucci, A. A. Saharian, H. G. Sargsyan, and V. V.
Vardanyan, Fermionic vacuum currents in topologically nontrivial
braneworlds: Two-brane geometry, Phys. Rev. D \textbf{101}, 045020 (2020).

\bibitem{Bell18} S. Bellucci, A. A. Saharian, D. H. Simonyan, and V.
Vardanyan, Fermionic currents in topologically nontrivial braneworlds, Phys.
Rev. D \textbf{98}, 085020 (2018).

\bibitem{Oliv19} W. Oliveira dos Santos, H. F. Mota, and E. R. Bezerra de
Mello, Induced current in high-dimensional AdS spacetime in the presence of
a cosmic string and a compactified extra dimension, Phys. Rev. D \textbf{99}%
, 045005 (2019); W. Oliveira dos Santos, E. R. Bezerra de Mello, and H. F.
Mota, Vacuum polarization in high-dimensional AdS space-time in the presence
of a cosmic string and a compactified extra dimension, Eur. Phys. J. Plus.
\textbf{135}, 27 (2020).

\bibitem{Bell20} S. Bellucci, W. Oliveira dos Santos, and E. R. Bezerra de
Mello, Induced fermionic current in AdS spacetime in the presence of a
cosmic string and a compactified dimension, Eur. Phys. J. C \textbf{80}, 963
(2020).

\bibitem{Bell21b} S. Bellucci, W. Oliveira dos Santos, E. R. Bezerra de
Mello, and A.A. Saharian, Topological effects in fermion condensate induced
by cosmic string and compactification on AdS bulk, Symmetry \textbf{14}, 584
(2022).

\bibitem{Bell22} S. Bellucci, W. Oliveira dos Santos, E. R. Bezerra de
Mello, and A.A. Saharian, Fermionic vacuum polarization around a cosmic
string in compactified AdS spacetime, JCAP01(2022)010.

\bibitem{Eliz94} E. Elizalde, S. Leseduarte, and S. D. Odintsov, Chiral
symmetry breaking in the Nambu-Jona-Lasinio model in curved spacetime with a
nontrivial topology, Phys. Rev. D \textbf{49}, 5551 (1994); E. Elizalde, S.
Leseduarte, and S. D. Odintsov, Higher derivative four-fermion model in
curved spacetime, Phys. Lett. B \textbf{347}, 33 (1995); D. K. Kim and G.
Koh, Restoration of dynamically broken chiral symmetry by a toroidal
space-time, Phys. Rev. D \textbf{51}, 4573 (1995).

\bibitem{Eliz13} E. Elizalde, S. D. Odintsov, and A. A. Saharian, Fermionic
Casimir densities in anti-de Sitter spacetime, Phys. Rev. D \textbf{87},
084003 (2013).

\bibitem{Abra} \textit{Handbook of Mathematical Functions}, edited by M.
Abramowitz and I.A. Stegun (Dover, New York, 1972).

\bibitem{Beze10} E. R. Bezerra de Mello, V. B. Bezerra, A. A. Saharian, and
V. M. Bardeghyan, Fermionic current densities induced by magnetic flux in a
conical space with a circular boundary, Phys. Rev. D \textbf{82}, 085033
(2010).

\bibitem{Grad} I. S. Gradshteyn and I. M. Ryzhik. \textit{Table of
Integrals, Series and Products} (Academic Press, New York, 1980).

\bibitem{Prud2} A. P. Prudnikov, Yu. A. Brychkov, and O. I. Marichev,\textit{%
\ Integrals and series}, Vol. 2 (Gordon and Breach, New York, 1986).

\bibitem{Knap04} A. A. Saharian and M. R. Setare, The Casimir effect on
background of conformally flat brane-world geometries, Phys. Lett. B \textbf{%
552}, 119 (2003); R. A. Knapman and D. J. Toms, Stress-energy tensor for a
quantized bulk scalar field in the Randall-Sundrum brane model, Phys. Rev. D
\textbf{69}, 044023 (2004); A. A. Saharian, Wightman function and Casimir
densities on AdS bulk with application to the Randall-Sundrum braneworld,
Nucl. Phys. B \textbf{712}, 196 (2005); A. S. Kotanjyan and A. A. Saharian,
Electromagnetic quantum effects in anti-de Sitter spacetime, Phys. At. Nucl.
\textbf{80}, 562 (2017); A. A. Saharian , A. S. Kotanjyan, and H. G.
Sargsyan, Electromagnetic field correlators and the Casimir effect for
planar boundaries in AdS spacetime with application in braneworlds, Phys.
Rev. D \textbf{102}, 105014 (2020).

\bibitem{Beze12b} E.~R.~Bezerra de Mello, A.~A.~Saharian, and S.~V.~Abajyan,
Fermionic vacuum polarization by a flat boundary in cosmic string spacetime,
Class. Quant. Grav. \textbf{30}, 015002 (2013).

\bibitem{Rand99b} L. Randall and R. Sundrum, An alternative to
compactification, Phys. Rev. Lett. \textbf{83}, 4690 (1999).

\bibitem{Flac01b} A. Flachi, I. G. Moss, and D. J. Toms, Fermion
stabilisation of extra dimensions and natural mass hierarchies, Phys. Rev. D
\textbf{64}, 105029 (2001).

\bibitem{Saha20} A. A. Saharian, Quantum vacuum effects in braneworlds on
AdS bulk, Universe \textbf{6}, 181 (2020).

\bibitem{Beze08cyl} E. R. Bezerra de Mello, V. B. Bezerra, A. A. Saharian,
and A. S. Tarloyan, Fermionic vacuum polarization by a cylindrical boundary
in the cosmic string spacetime, Phys. Rev. D \textbf{78}, 105007 (2008).

\bibitem{Frol87} V. P. Frolov and E. M. Serebriany, Vacuum polarization in
the gravitational field of a cosmic string, Phys. Rev. D \textbf{35}, 3779
(1987).

\bibitem{Saha07} A. A. Saharian, The generalized Abel-Plana formula with
applications to Bessel functions and Casimir effect, arXiv:0708.1187
[hep-th].
\end{thebibliography}
\end{document}